\documentclass{article}

\usepackage{latexsym, amsmath, amssymb, mathrsfs, epsfig, graphicx, color}  
\numberwithin{equation}{section}

\graphicspath{{eps/}}

\newcommand{\myscript}[2]{\begin{array}{c}
    {\scriptstyle #1} \\
    {\scriptstyle #2}
  \end{array}}

\newcommand{\af}{antiferromagnetic}
\newcommand{\dbar}{{\partial\!\!\!/}}
\newcommand{\Deltab}{\bar{\Delta}}
\newcommand{\Ec}{\mathcal{E}}
\newcommand{\et}{\widetilde{e}}
\newcommand{\half}{\frac{1}{2}}
\newcommand{\halfpi}{\frac{\pi}{2}}
\newcommand{\im}{{\rm Im}\ }
\newcommand{\Jhat}{\widehat{J}}
\newcommand{\Khat}{\widehat{K}}
\newcommand{\kt}{\widetilde{k}}
\newcommand{\mt}{\widetilde{m}}
\newcommand{\qb}{\bar{q}}
\newcommand{\Rc}{\check{R}}
\newcommand{\RRc}{\check{\cal R}}
\newcommand{\rhohat}{\widehat{\rho}}
\newcommand{\rme}{{\rm e}}
\newcommand{\rmi}{{\rm i}}
\newcommand{\Sc}{\mathcal{S}}
\newcommand{\SU}{{\rm SU}}
\newcommand{\Uq}{{\rm U}_q({\rm sl}_2)}
\newcommand{\Zbb}{\mathbb{Z}}
\newcommand{\Zc}{\mathcal{Z}}
\newcommand{\Zhat}{\widehat{Z}}
\newcommand{\til}{\widetilde}

\title{The $\Zbb_2$ staggered vertex model and its applications}
\author{Yacine Ikhlef$^{1}$, Jesper Lykke Jacobsen$^{2,3}$, and Hubert Saleur$^{3,4}$ \\ [2.0mm]
  ${}^1$Section de math\'ematiques, Universit\'e de Gen\`eve, \\
  2-4 rue du Li\`evre, CP 64, 1211 Gen\`eve 4, Suisse\\
  ${}^2$LPTENS, 24 rue Lhomond, 75231 Paris, France \\
  ${}^3$Institut de Physique Th\'eorique, CEA Saclay,
  91191 Gif Sur Yvette, France \\
  ${}^4$Department of Physics,
  University of Southern California, Los Angeles, CA 90089-0484}

\begin{document}

\maketitle

\begin{abstract}
  New solvable vertex models can be easily obtained by staggering the spectral
  parameter in already known ones. This simple construction reveals some
  surprises: for appropriate values of the staggering, highly non-trivial
  continuum limits can be obtained. The simplest case of staggering with
  period two (the $\Zbb_2$ case) for the six-vertex model was shown to be
  related, in one regime of the spectral parameter, to the critical
  antiferromagnetic Potts model on the square lattice, and has a non-compact
  continuum limit. Here, we study the other regime: in the very anisotropic
  limit, it can be viewed as a zig-zag spin chain with spin anisotropy, or as
  an anyonic chain with a generic (non-integer) number of species. From the
  Bethe-Ansatz solution, we obtain the central charge $c=2$, the conformal
  spectrum, and the continuum partition function, corresponding to one free
  boson and two Majorana fermions.  Finally, we obtain a massive integrable
  deformation of the model on the lattice. Interestingly, its scattering
  theory is a massive version of the one for the flow between minimal models.
  The corresponding field theory is argued to be a complex version of the
  $C_2^{(2)}$ Toda theory.
\end{abstract}

\section{Introduction}
\label{sec:intro}

\begin{figure}
  \begin{center}
    \includegraphics{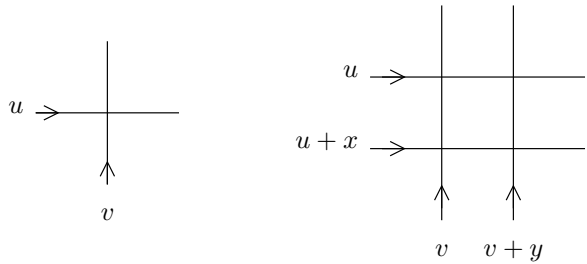}
  \end{center}
  \caption{The $\Rc$-matrices for a basic vertex model (left), and for a
    staggered vertex model with $n=2$ (right).}
  \label{fig:R-stag}
\end{figure}

It is a simple consequence of the quantum inverse scattering~\cite{IKB}
formalism (going back to Baxter's `Z invariance'~\cite{Baxter}) that new
integrable vertex models can be obtained from basic ones by allowing for some
staggering of the spectral parameters. If the basic $\Rc$-matrix is associated
with a single crossing , one can in this way build `block' $\Rc$-matrices,
using $n^2$ crossings, with $n=2,3,\ldots$, and staggering the spectral
parameters (see Fig.~\ref{fig:R-stag}).

While constructing the model and writing down the Bethe equations is
straightforward, the physics of the models thus obtained presents interesting
subtleties.  A striking example was discussed in detail in
Refs.~\cite{JS-PAF,IJS-PAF}, in relation with the {\af} Potts
model~\cite{Baxter-PAF}. It was found then that the case $n=2$ for the
six-vertex model has a non-compact continuum limit~\cite{IJS-PAF} in a certain
regime of the spectral parameter (see below for a more accurate definition),
and may be related to the complex sine-Gordon (SG) model.  While major
difficulties remain in this case, the other regime of the spectral parameter
turns out to be also of interest, and somewhat more tractable. Its study is
the main goal of this paper.

Before launching into details, a general discussion about vertex models, spin
chains and field theories seems useful. Indeed, the correspondence between
integrable spin chains with $\SU(2)$ or $\SU(2)_q$ symmetry and quantum field
theories has been investigated in great details already. In the {\af} regime,
it is well known that a chain of spin $s$ corresponds to a level-$2s$
Weiss-Zumino-Witten (WZW) model in the $\SU(2)$ case, and to a deformation of
this theory in the Cartan direction for $\SU(2)_q$~\cite{Alcaraz}. Cursory
examination of the literature would suggest nothing much remains to be done in
this area.

From the $R$-matrix point of view, models with higher values of the spin $s$
are obtained by projecting tensor products of ($2s$) fundamental
representations of $\Uq$ onto higher spin. This construction has a close
parallel in conformal field theory (CFT), where higher level ($2s$)
representations of current algebra are obtained by combining ($2s$) level one
representations. There are several reasons why it would be interesting to
build integrable models which are {\it not} projected onto irreducible $\Uq$
components. In the case $s=1$ for instance, this would correspond to models of
pairs of spin-$\half$ variables. This can be reinterpreted more physically in
terms of ladders, or, in terms of electron physics, pairs of wires or
channels. The latter case is of the highest importance. For instance, the
two-channel Kondo model~\cite{AffleckLudwig} or the two-channel Interacting
Resonant Level models~\cite{Natan,Edouard} are usually solved by going to an
even-odd basis, which effectively amounts to solving the problem in the level
$k=2$ sector. Many physical questions are however related to the mixture of
the even and odd degrees of freedom -- {\it e.g.} because it corresponds to
transport of electrons between wires.  The search for integrable cases where
this mixture could be studied is a priority.

To have a better idea of what to expect, it is useful to turn to the CFT point
of view. Imagine starting with two $\SU(2)$ Kac-Moody algebras at level 1,
represented by the currents $j_i^\mu$, $i=1,2$ and $\mu=1,2,3$. The sums
\begin{equation}
  J^\mu =j_1^\mu + j_2^\mu
\end{equation}
are well known to provide then a Kac-Moody algebra at level 2. Of course, each
level 1 corresponds to central charge $c=1$, while level 2 has central charge
$c=3/2$. The point is that in taking two copies of level 1, an Ising model CFT
factors out, according to the well known-decomposition
\begin{equation}
  \SU(2)_1 \times \SU(2)_1 = \SU_2(2) + \hbox{ Ising} \,.
\end{equation}
This can be illustrated quickly using bosonisation. Introduce two chiral
bosons $\phi_1,\phi_2$ with propagators
\begin{equation}
  \langle \phi_i(z)\phi_i(w) \rangle = -\frac{1}{4\pi} \ln(z-w) \,.
\end{equation}
The two level-1 current algebras are obtained through
\begin{equation}
  j_i^\pm \propto \exp \left(\pm \rmi \sqrt{8\pi}\phi_i \right) \,,
  \qquad j_i^3 \propto \partial\phi_i \,.
\end{equation}
It is convenient to introduce now symmetric and antisymmetric combinations of
the bosons:
\begin{equation}
  \Phi = \frac{1}{\sqrt{2}} \left( \phi_1+\phi_2 \right) \,,
  \qquad \phi = \frac{1}{\sqrt{2}} \left( \phi_1-\phi_2 \right) \,.
\end{equation}
So we have
\begin{equation}
  J^\pm \propto \cos\sqrt{4\pi}\phi \ \exp(\pm \rmi\sqrt{4\pi}\Phi) \,, \\
  \qquad J^3 \propto \partial \Phi \,.
\end{equation}
The field $\cos(\sqrt{4\pi}\Phi)$ is a Majorana fermion~\cite{Kiritsis}. The
field $\rmi\sin(\sqrt{4\pi}\phi)$ is another one, which is orthogonal to the
currents $J^\mu$, and is discarded in the construction of $\SU(2)_2$.
Corresponding to this splitting, the sum $H=H_1+H_2$ of the two one-boson
Hamiltonians decomposes as $H= H_{\SU(2)_2} +H_{\rm Ising}$ where
$H_{\SU(2)_2} \propto \sum_{\mu} :J^\mu J^\mu:$, and $J^\mu$ are the currents
at level two.

Quantum deformations of $\SU(2)_2$ are obtained by adding to the Hamiltonian
$H_{\SU(2)_2}$ a Cartan deformation proportional to $J^3J^3$. We can as well
deform the full Hamiltonian, obtaining in this way a theory made of two
Majorana fermions and one boson with anisotropy-dependent radius. This should
be the continuum limit of the models we are after. We will show in the
following that these models are obtained by the general staggering
construction, with $n=2$ and an appropriate choice of the spectral parameters.

Interestingly, it can be shown~\cite{Frahm} that the staggered models
correspond algebraically to solutions of the Yang-Baxter equations based on
`bigger' irreducible representations of the quantum affine algebra ${\rm
  U}_q(\widehat{\rm sl}_2)$. This occurs ultimately because finite-dimensional
irreducible representations of quantum affine algebras are isomorphic to
products of evaluation representations, which are themselves `decorations'
(with the spectral parameter) of the usual spin-$s$ representations of
$\Uq$~\cite{Chari}.

Another important application of the staggered model is that it can be used to
produce a lattice discretisation of a {\it massive} QFT. This is done,
following Ref.~\cite{RS}, by introducing into the staggered model an
additional (purely imaginary) staggering of the spectral parameters. Using
this approach, we obtain a scattering theory involving two types of massive
particles, where the scattering between particles of the same type
(resp. different types) is given by the Sine-Gordon $S$-matrix (resp. the
Sine-Gordon $S$-matrix with an imaginary shift in the rapidity). It turns out
that such a scattering theory (but with massless particles) arose
before~\cite{FSZ} in the context of minimal models of CFT perturbed by the
$\Phi_{13}$ operator. Consider the action:
\begin{equation} \label{eq:MAp}
  A = A_{\rm min} + \lambda \int d^2x \, \Phi_{13}(x) \,,
\end{equation}
where $A_{\rm min}$ is the action of a minimal model of CFT.  The problem of
finding the Thermodynamic Bethe Ansatz (TBA) equations for the
renormalisation-group flow of the theory~\eqref{eq:MAp} was first studied by
Zamolodchikov in
Ref.~\cite{TBA-RSOS,TBA-massless}.  The results depend crucially on the sign
of the coupling $\lambda$.  For $\lambda<0$, the model~\eqref{eq:MAp} becomes
massive. It was shown in Ref.~\cite{TBA-RSOS} that the corresponding
$S$-matrix is the simple RSOS $S$-matrix and that the TBA diagram is of the
$A_n$ type, with a massive particle at one end of the diagram. For
$\lambda>0$, the model~\eqref{eq:MAp} describes the massless flow between two
consecutive minimal models. In Ref.~\cite{TBA-massless}, a TBA diagram was
proposed, without resorting to an $S$-matrix: this diagram is also of the
$A_n$ type, but with mass terms $\rme^{\pm \theta}$ at the two ends of the
diagram.  It was found later, in Ref.~\cite{FSZ}, that the corresponding
scattering theory consists in massless left/right particles, interacting
through the SG and shifted-SG $S$-matrices.

In summary, we construct here a non-critical lattice model whose continuum
excitations are described by a massive version of the $S$-matrix for the flow
between minimal models of CFT. We also propose an effective QFT for this
model, with one boson and two Majorana fermions which interact with each
other.

The paper is organised as follows. In Section~\ref{sec:potts}, we expose in
more detail the construction of the model and its various lattice
formulations, as well as the relation with spin chains and anyonic chains.  In
Section~\ref{sec:bethe-ansatz}, we present the Bethe-Ansatz solution, and
obtain the critical exponents, through the study of low-energy excitations.
In Section~\ref{sec:Z-torus}, we discuss the associated CFT, and exhibit the
full operator content through the study of torus partition functions.
Finally, in Section~\ref{sec:massive}, we study the integrable massive
deformation, and derive its scattering theory, TBA equations and ground-state
energy scaling function. This allows us to propose an interacting effective
QFT.  Some important but quite long calculations are done in the Appendices.

\section{Solvable Potts models and loop Hamiltonians}
\label{sec:potts}

In this section, we first recall the definition of the Potts model~\cite{Potts} on the square lattice
and its equivalence to a loop model based on the Temperley-Lieb (TL) algebra~\cite{TL}.
Then we introduce a solvable case, involving a staggering of the spectral
parameters, and we obtain the expression~\eqref{eq:H} for the associated
one-dimensional Hamiltonian, in terms of the TL generators.
This solvable model has two critical regimes: regime I corresponds to the {\af} critical
transition~\cite{JS-PAF,IJS-PAF,Baxter-PAF}; regime II contains a mixture of
ferromagnetic and {\af} couplings, and is the subject of the present
paper. Finally, we explain the relation to Majumdar-Ghosh spin chains~\cite{MG,qMG} and anyonic
chains~\cite{anyons1, anyons2}.

\subsection{The Potts model and the Temperley-Lieb algebra}

\begin{figure}
  \begin{center}
    \includegraphics{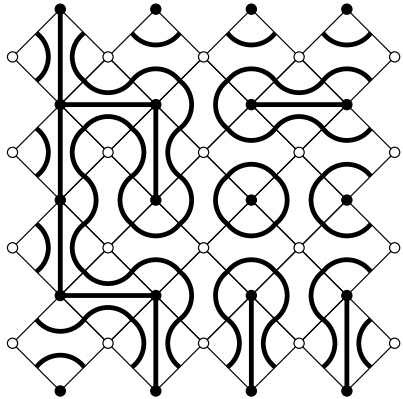}
    \caption{Equivalence between the cluster and loop models. The full dots represent Potts spins, and the
empty dots are the sites of the dual lattice. The thin lines are the edges of the lattice $\cal L$. Each cluster configuration corresponds to precisely
one loop configuration.}
    \label{fig:FK-TL}
  \end{center}
\end{figure}

The $Q$-state Potts model~\cite{Potts} is a model of classical spins with nearest-neighbour interactions.
Each spin $S_j$ can take $Q$ values, and sits on a vertex of the square lattice. The Boltzmann
weights and the partition function are given by:
\begin{equation}
  \begin{array}{rcl}
    W[\{ S_j \}] &=& {\displaystyle \prod_{\langle ij \rangle} \exp \left[ J_{ij} \ \delta(S_i,S_j) \right]} \,, \\
    Z_{\rm Potts} &=& {\displaystyle \sum_{\{ S_j \}} W[\{ S_j \}]} \,.
  \end{array}
\end{equation}
where the product runs over all the bonds of the lattice, and the $J_{ij}$ are the coupling constants of the
model. This model can be reformulated as a cluster model, called the Fortuin-Kasteleyn model~\cite{FK},
in the following way. We may write the Boltzmann weights as:
\begin{equation}
  W[\{ S_j \}] = \prod_{\langle ij \rangle} \left[ 1 + v_{ij} \delta(S_i,S_j) \right] \,,
  \qquad v_{ij}\equiv e^{J_{ij}}-1 \,.
\end{equation}
The above expression can be expanded, and each term in the expansion is represented by a subgraph of the
square lattice, consisting of the bonds where the $\delta$ term has been chosen. When we sum over the spin
configurations, the subgraph gets a weight $Q$ for each of its connected components.
So the partition function can be written as:
\begin{equation}
  Z_{\rm FK} = \sum_{G} Q^{C(G)} \prod_{\langle ij \rangle \in G} v_{ij} \,,
\end{equation}
where the sum is over all possible subgraphs (or cluster configurations) $G$, and $C(G)$ is the number of
connected components of $G$. The Fortuin-Kasteleyn model is, in turn, transformed to a dense loop model.
Consider the lattice $\cal L$ obtained by the union of the original square lattice and its dual. Each face
of $\cal L$ contains exactly one Potts bond. By decorating these faces as shown in
Fig.~\ref{fig:FK-TL}, we obtain configurations of closed loops on the lattice.
The partition function is then:
\begin{equation}
  Z_{\rm loop} = \sum_{G} Q^{L(G)/2}  \prod_{\langle ij \rangle \in G} \frac{v_{ij}}{\sqrt{Q}} \,,
\end{equation}
where $L(G)$ is the number of closed loops.

\begin{figure}
  \begin{center}
    \includegraphics{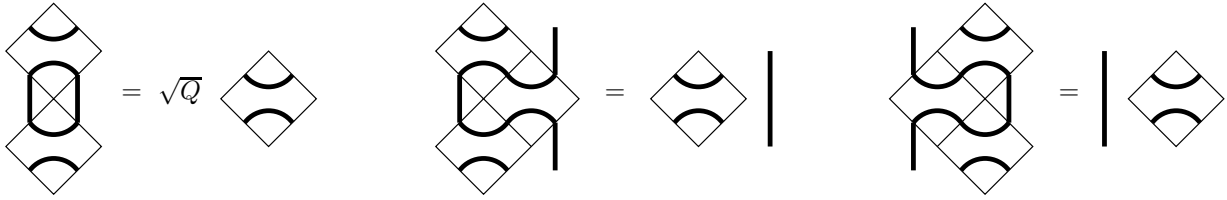}
    \caption{The Temperley-Lieb algebraic relations~\eqref{eq:TL-def}.}
    \label{fig:TL-def}
  \end{center}
\end{figure}

In the present study, we will restrict ourselves to the case when the couplings are $J_{ij}=J_1$
on horizontal bonds and $J_{ij}=J_2$ on vertical bonds. Then, for the loops, a plaquette can be of
two types, according to the direction of the Potts bond it contains. The local Boltzmann weights for
the loop plaquettes can be written:
\begin{equation} \label{eq:plaq-tl}
  \includegraphics{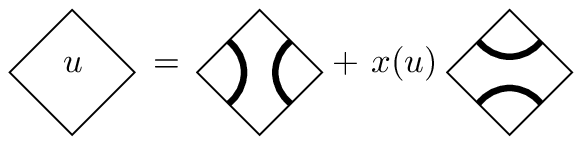}
\end{equation}
where $x(u)=v_1/\sqrt{Q}$ for plaquettes of type 1 and $x(u)=\sqrt{Q}/v_2$ for plaquettes of type 2.
The non-local weights for closed loops are encoded in the Temperley-Lieb algebra~\cite{TL}, with generators
$(e_j)_{j=1 \dots 2N}$:
\begin{equation} \label{eq:TL-def}
  \begin{array}{l}
    e_j^2 = \sqrt{Q} \ e_j \\
    e_j e_{j\pm 1} e_j = e_j \\
    e_j e_{j'} = e_{j'} e_j \qquad \text{if $|j-j'|>1$} \,.
  \end{array}
\end{equation}
These relations are interpreted graphically as follows: $j$ is the site index in the horizontal direction,
and $e_j$ is the represented by the second term of Eq.~\eqref{eq:plaq-tl} at position $j$
(see Fig.~\ref{fig:TL-def}). The loop weight $\sqrt{Q}$ is parameterised as:
\begin{equation}
  \sqrt{Q} = 2 \cos \gamma \,, \qquad 0 \leq \gamma \leq \halfpi \,.
\end{equation}

\subsection{Solvable inhomogeneous Potts model}

To construct a solvable model, the first step is to obtain an $\Rc$-matrix which satisfies the
Yang-Baxter equations:
\begin{equation} \label{eq:YBE}
  \Rc_{j,j+1}(u) \Rc_{j+1,j+2}(u-v) \Rc_{j,j+1}(v) = \Rc_{j+1,j+2}(v) \Rc_{j,j+1}(u-v) \Rc_{j+1,j+2}(u) \,.
\end{equation}
The Temperley-Lieb algebra~\eqref{eq:TL-def} provides a solution to these equations:
\begin{equation} \label{eq:Rmat}
  \Rc_{j,j+1}(u) \equiv \sin(\gamma-u) \ 1 + \sin u \ e_j \,.
\end{equation}
Comparing with~\eqref{eq:plaq-tl}, the local weight $x$ is given by:
\begin{equation}
  x(u) = \frac{\sin u}{\sin(\gamma-u)} \,.
\end{equation}
Using this $\Rc$-matrix, we can construct a solvable model on the square lattice, by choosing 
the spectral parameters along the lines of the lattice. Suppose, from now on, that the square
lattice where the loops live consists of horizontal and vertical lines. To respect the alternation
$J_1,J_2$ of the Potts coupling constants, we have to use the spectral parameters
$(u,u+\alpha,u,u+\alpha,\dots)$ and $(0,\alpha,0,\alpha,\dots)$, and also ensure that
$\Rc(u+\alpha) \propto \Rc(u-\alpha)$.
This holds only for $\alpha=0$ or $\alpha=\pi/2$. 
\begin{figure}
  \begin{center}
    \includegraphics{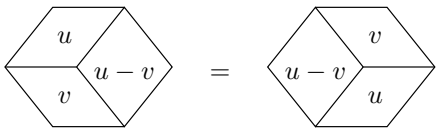}
    \caption{The Yang-Baxter equations~\eqref{eq:YBE}.}
    \label{fig:YBE}
  \end{center}
\end{figure}

The case $\alpha=0$ is the homogeneous TL loop model, and corresponds to the
well-studied self-dual Potts model. It has two critical regimes: $0<u<\gamma$
is the ferromagnetic critical transition, and $\gamma<u<\pi$ is the `non-physical self-dual line',
governing the critical Berker-Kadanoff phase \cite{SaleurPotts}.

\begin{figure}
  \begin{center}
    \includegraphics{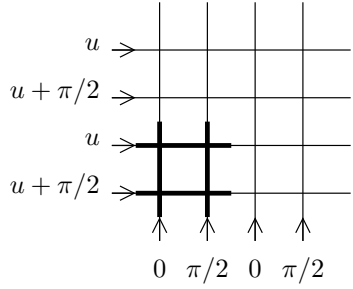}
    \caption{Lines of spectral parameters for the staggered model. The block $\RRc$-matrix is represented
      in bold lines.}
    \label{fig:block-Rmat}
  \end{center}
\end{figure}

In the present paper, we are interested in the case $\alpha=\pi/2$, which we
call the $\Zbb_2 $ staggered model.
The parameters $v_1,v_2$ are then given by:
\begin{equation}
  \frac{v_1}{\sqrt{Q}} = \frac{\sin u}{\sin(\gamma-u)} \,,
  \qquad \frac{v_2}{\sqrt{Q}} = -\frac{\cos(\gamma-u)}{\cos u} \,.
\end{equation}
There are again two regimes:
\begin{itemize}
  \item Regime I: $\gamma<u<\pi/2$. \\
    In this regime, we have $v_1<0$ and $v_2<0$. This was identified by Baxter~\cite{Baxter-PAF}
    as the location of the {\af} critical transition. 
    There is an isotropic point $u=\frac{\pi}{2}+\frac{\gamma}{2}$ for which $v_1=v_2$.
    The exact solution on
    the lattice, and the corresponding
    field theory were studied in detail previously~\cite{JS-PAF,IJS-PAF}.
  \item Regime II: $0<u<\gamma$. \\
    In this regime, we have now $v_1>0$ and $v_2<0$.
    There is thus no isotropic point, but for $u=\frac{\gamma}{2}$ we have that $v_1=-v_2$ differ only by a sign; we shall see that at this point isotropy is recovered in the continuum limit. 
    In the following, we will focus our attention on
    this regime.
\end{itemize}
Let us now recall the lattice structure of the staggered model~\cite{JS-PAF,IJS-PAF}.
This is better described by the block $\RRc$-matrix (see Fig.~\ref{fig:block-Rmat}):
\begin{equation}
  \RRc_{j,j+1}(u) \equiv \Rc_{2j,2j+1}(u-\pi/2) \Rc_{2j-1,2j}(u) \Rc_{2j+1,2j+2}(u) \Rc_{2j,2j+1}(u+\pi/2) \,.
\end{equation}
Using~\eqref{eq:Rmat}, we get the expression for $\RRc_{j,j+1}(u)$ in terms of the TL generators $e_j$:
\begin{eqnarray}
  \RRc_{j,j+1}(u)  &=& -\frac{1}{4} \sin^2(2\gamma-2u) \ 1
  - \half \sin u \sin (2\gamma-2u) \left[ \cos(\gamma-u) (e_{2j-1}+e_{2j+1}) 
    + 2\cos \gamma \cos u \ e_{2j} \right] \notag \\
  && + \frac{1}{4} \sin 2u \sin(2\gamma-2u) \ (e_{2j-1}e_{2j}+e_{2j}e_{2j-1} + e_{2j}e_{2j+1}+e_{2j+1}e_{2j}) \notag \\
  && + \sin^2 u \cos u \left[ \cos(\gamma-u) \ (e_{2j-1} e_{2j+1} e_{2j} + e_{2j} e_{2j-1} e_{2j+1}) 
    - \cos u \ e_{2j} e_{2j-1} e_{2j+1} e_{2j} \right] \,. \label{eq:block-Rmat}
\end{eqnarray}
As a consequence of the Yang-Baxter equations, $\RRc_{j,j+1}(u)$ commutes with the operator:
\begin{equation}
  c_j \equiv (\cos \gamma)^{-2} \ \Rc_{2j-1,2j}(\pi/2) \Rc_{2j+1,2j+2}(\pi/2) \,.
\end{equation}
Furthermore, using the TL algebraic relations, we see that $c_j^2=1$. Thus,
{\bf the $\RRc$-matrix has a $\Zbb/2\Zbb$ symmetry}, arising from the
staggered structure. We call $t(u)$ the one-row transfer matrix, with
horizontal spectral parameter $u$ and vertical spectral parameters $(0,\pi/2,0,\pi/2,\dots)$. 
We consider a lattice of width $2N$ sites and height $2M$ sites.
The partition function of the staggered model with periodic boundary conditions is then:
\begin{equation}
  Z_{MN} = {\rm Tr} \ \left[ t(u) t(u+\pi/2) \right]^M \,.
\end{equation}
The two-row transfer-matrix $t(u) t(u+\pi/2)$ commutes with the $\Zbb/2\Zbb$
`charge operator':
\begin{equation}
  C \equiv \prod_{j=1}^N \left[ (\cos \gamma)^{-1} \ \Rc_{2j-1,2j}(\pi/2) \right] \,.
\end{equation}

\subsection{Very anisotropic limit}

An interesting aspect of Yang-Baxter integrable statistical models is that the
transfer matrix generally possesses a {\it very anisotropic limit}, where its
derivatives with respect to $u$ are local, one-dimensional Hamiltonians. In
the expression~\eqref{eq:block-Rmat} for the block $\RRc$-matrix, we observe
that $\RRc_{j,j+1}(0) = -\frac{1}{4} \sin^2 2\gamma \ 1$, and so the two-row
transfer-matrix reduces to a cyclic translation of two sites to the right in
the limit $u \to 0$. The first-order Hamiltonian is:
\begin{equation}
  H \equiv -\frac{1}{2} \sin 2\gamma 
  \ \left. \frac{{\rm d} \log[t(u)t(u+\pi/2)]}{{\rm d} u} \right|_{u=0}
  = -\frac{1}{2} \sin 2\gamma 
  \ \sum_{j=1}^N \RRc_{j,j+1}(0)^{-1} \frac{{\rm d} \RRc_{j,j+1}}{{\rm d}u}(0) \,,
\end{equation}
which gives:
\begin{equation} \label{eq:H} H = 2N \cos 2\gamma \ 1 + \sum_{j=1}^{2N} \left(
    -2 \cos \gamma \ e_j + e_je_{j+1} + e_{j+1}e_j \right) \,.
\end{equation}

\begin{figure}
  \begin{center}
    \includegraphics{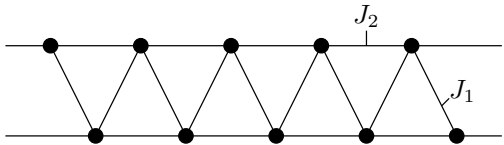}
    \caption{Zig-zag spin chain with $J_1$ and $J_2$ interactions.}
    \label{fig:zigzag}
  \end{center}
\end{figure}

For a generic value of $\gamma$, the TL algebra can be represented as acting
on a chain of spin-$\half$ variables with $\Uq$ symmetry (for $q=\rme^{\rmi
  \gamma}$). Indeed, consider the Hilbert space $[{\rm span}(|\uparrow\rangle,
|\downarrow \rangle)]^{\otimes 2N}$, describing $2N$ spins
$(\sigma_1,\dots,\sigma_{2N})$, and take two consecutive spins
$\sigma_j,\sigma_{j+1}$: the total spin $\sigma_j+\sigma_{j+1}$ can be in the
representation of spin $0$ or $1$.  We call $P^{(0)}_{j,j+1}$ the projector
onto spin-$0$ according to this decomposition. Then it turns out that the
unnormalised projectors
\begin{equation} \label{eq:ej-Pj}
  e_j \equiv 2\cos \gamma \ P^{(0)}_{j,j+1}
\end{equation}
satisfy the TL algebra~\eqref{eq:TL-def}.  In terms of the Pauli matrices,
the generators~\eqref{eq:ej-Pj} can be written:
\begin{equation} \label{eq:ej-sigma}
  e_j = -(\sigma_j^+ \sigma_{j+1}^- +
  \sigma_j^- \sigma_{j+1}^+) +\half (1-\sigma_j^z \sigma_{j+1}^z) \rme^{\rmi
    \gamma \sigma_{j+1}^z} \,.
\end{equation}
Using this representation, the first summand in~\eqref{eq:H} would simply give
the XXZ spin chain. The quadratic terms in $e_j,e_{j+1}$ are due to the
staggering, and lead to a different spin model.  After some algebra with the
Pauli matrices, we obtain:
\begin{eqnarray} \label{eq:ej2-sigma} e_j e_{j+1} + e_{j+1} e_j &=& \half +
  (\sigma_j^+ \sigma_{j+2}^- + \sigma_j^- \sigma_{j+2}^+)
  + \half \sigma_j^z \sigma_{j+2}^z \nonumber \\
  && - \rme^{\rmi \gamma \sigma_{j+2}^z} (\sigma_j^+ \sigma_{j+1}^- +
  \sigma_j^- \sigma_{j+1}^+)
  - \half \sigma_j^z \sigma_{j+1}^z \nonumber \\
  && - \rme^{-\rmi \gamma \sigma_j^z} (\sigma_{j+1}^+ \sigma_{j+2}^- +
  \sigma_{j+1}^- \sigma_{j+2}^+) - \half \sigma_{j+1}^z \sigma_{j+2}^z \,.
\end{eqnarray}
Let us write our quadratic TL Hamiltonian~\eqref{eq:H} more generally as:
\begin{equation} \label{eq:H-TL} H(K_1,K_2) = K_1 \sum_{j=1}^{2N} e_j + K_2
  \sum_{j=1}^{2N} (e_j e_{j+1} + e_{j+1} e_j) \,.
\end{equation}
Then, from~\eqref{eq:ej-sigma}-\eqref{eq:ej2-sigma}, the above Hamiltonian can
be written:
\begin{equation} \label{eq:H-spin} H(K_1,K_2) = \sum_{j=1}^{2N} \left\{
    \frac{J_2}{2} \boldsymbol{\sigma}_j \cdot \boldsymbol{\sigma}_{j+2} +
    J_1^{xy} (\sigma_j^+\sigma_{j+1}^- + \sigma_j^-\sigma_{j+1}^+) + J_1^z
    \sigma_j^z \sigma_{j+1}^z + \rmi J_3
    (\sigma_{j-1}^z-\sigma_{j+2}^z)(\sigma_j^+\sigma_{j+1}^- +
    \sigma_j^-\sigma_{j+1}^+) \right\} \,,
\end{equation}
where:
\begin{equation}
  J_1^{xy}= -(K_1 + 2\cos\gamma \, K_2) \,,
  \quad J_1^z = - \left(\half \cos \gamma \, K_1 + K_2 \right) \,,
  \qquad J_2=K_2 \,,
  \qquad J_3=K_2 \sin \gamma \,.
\end{equation}
The $J_2$ term in equation~\eqref{eq:H-spin} represents two XXX spin chains,
living on the even and odd sites respectively. The $J_1$ terms correspond to
an XXZ interaction with a `zig-zag' shape (see Fig.~\ref{fig:zigzag}).  The
$J_3$ term is an anti-Hermitian three-spin interaction, with no obvious
physical interpretation.

In the case of our $\Zbb_2$ staggered model, we have $(K_1, K_2) = (-2\cos
\gamma, 1)$, and so $J_2=1, J_1^{xy}=0, J_1^z=-\sin^2 \gamma$.  Thus the
Hermitian terms of~\eqref{eq:H-spin} correspond to two {\af} XXX spin chains
with a ferromagnetic $-\sigma^z \sigma^z$ zig-zag interaction.  Let us discuss
the consequences of a naive bosonisation~\cite{bosonisation} argument for this
model.  We can discard the irrelevant $J_3$ term, and we end up with two free
bosons $\phi_1,\phi_2$ with compactification radii $R_1=R_2=R$ independent
from $\gamma$, coupled by the quadratic term $J_1^z \partial \phi_1 \partial
\phi_2$. Since this term is symmetric in the exchange of the bosons ($\phi_1
\leftrightarrow \phi_2$), the symmetric and antisymmetric combinations $\Phi =
(\phi_1 + \phi_2)/\sqrt{2}, \phi = (\phi_1 - \phi_2)/\sqrt{2}$ are decoupled
free bosons, with compactification radii $R_\pm = R \pm \delta R$, where
$\delta R$ depends on $\gamma$ through $J_1^z$.  Thus, we obtain two decoupled
free bosons $\Phi,\phi$ with both radii depending on $\gamma$. However, using
the Bethe Ansatz exact solution of the staggered model (see
Section~\ref{sec:bethe-ansatz}), we find that the continuum limit consists of
two free bosons with one radius depending on $\gamma$, and the other radius
independent of $\gamma$. It seems then that the bosonisation approach misses
an important effect due to the anti-Hermitian $J_3$ term in~\eqref{eq:H-spin}.

Let us go back to the model~\eqref{eq:H-TL} with general values of $K_1,K_2$,
which was studied numerically in Ref.~\cite{IJS-TL}. Additionally to the $\Zbb_2$
staggered integrable model, this Hamiltonian contains a remarkable point at
$(K_1, K_2)=(-4\cos \gamma, 1)$. This was identified in Ref.~\cite{qMG} as the
$q$-deformed version of the Majumdar-Ghosh (MG) spin chain~\cite{MG}.  The
latter was defined as an $\SU(2)$ spin-$\half$ chain with nearest- and
next-nearest-neighbour interactions, for which the totally dimerised state
(where the pairs of spins $\sigma_{2j-1},\sigma_{2j}$ form a singlet) is the
exact ground state. The excitations of the isotropic MG chain are known to be
gapped spinons~\cite{MG-spinons}. In Ref.~\cite{IJS-TL} it was proved that for
$\gamma=\pi/t$ with $t \geq 5$ integer, the dimerised state is still a ground
state, and numerical evidence was given for the existence of a gapped MG phase
in the model~\eqref{eq:H-TL} for generic $\gamma$ in the range $(n^* \simeq
1.5)<2\cos \gamma<2$.

\subsection{Relation to anyonic chains}

We end this Section by mentioning an additional motivation for the study of TL
Hamiltonians such as~\eqref{eq:H-TL}, which is their connection to `anyonic
chains', an object introduced and studied recently~\cite{anyons1,anyons2}.
The starting point is the Read-Rezayi construction~\cite{read-rezayi} of trial
wavefunctions for the Fractional Quantum Hall Effect. Following this approach,
the trial wavefunctions are given by the correlation functions of the $\Zbb_k$
parafermionic CFTs. The elementary excitations above the ground state ({\it
  anyons}) then obey the fusion rules given by these CFTs.

\begin{figure}
  \begin{center}
    \includegraphics{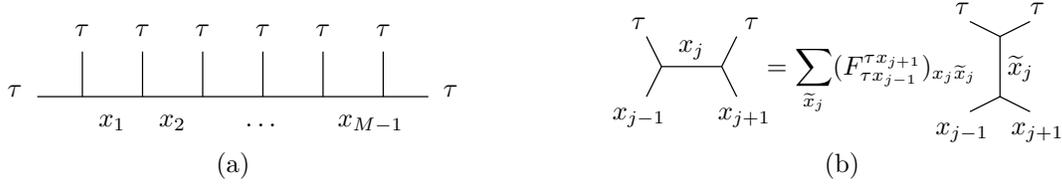}
    \caption{(a) Anyonic chain. (b) Transformation of fusion diagrams through
      the $F$-matrix. Figures reproduced from Ref.~\cite{anyons1}.}
    \label{fig:anyons}
  \end{center}
\end{figure}

The anyonic chain, defined in Ref.~\cite{anyons1} for the $\Zbb_3$
parafermionic CFT, consists of $M+2$ anyons with short-range interaction. Each
anyon lives in a Hilbert space whose basis elements are labelled by the
primary fields of the CFT. In the case of the $\Zbb_3$ parafermionic CFT, the
primary fields are $1$ and $\tau$, with the fusion rule:
\begin{equation}
  \tau \times \tau  = 1 + \tau \,.
\end{equation}
We fix $M+2$ anyons along a chain to be in the non-trivial state $\tau$, and
give an interaction energy to the different ways they can fuse with one
another. Suppose the $\tau$-anyons are allowed to fuse according to the linear
diagram shown in Fig.~\ref{fig:anyons}a, where the $x_j$ are also
anyons. Locally, the fusion diagram between two consecutive $\tau$-anyons and
the anyons $x_{j-1}, x_{j+1}$ can be rewritten in terms of a new anyonic
variable $\til{x}_j$, using the $F$-matrix (see
Fig.~\ref{fig:anyons}b). The variable $\til{x}_j$ now results from the
direct fusion of the two consecutive $\tau$-anyons.  We give energy
$-\epsilon$ to the configuration $\til{x}_j=1$, and energy $0$ to the
configuration $\til{x}_j=\tau$.  Let us describe the Hilbert space and
the interaction Hamiltonian in terms of the intermediary states $x_j$. The
basis of the Hilbert space is labelled by the words $x_1 \dots x_{M-1}$ on the
alphabet $\{1,\tau\}$ which are allowed by the fusion rules. These are the
words which do not have the word $11$ as a subword. Fig.~\ref{fig:anyons}b
defines a local change of basis $|x_j\rangle \to |\til{x}_j\rangle$. In
this new basis, the interaction term defined above is proportional to the
projector $\til{p}_j = |1\rangle \langle 1|$. Transforming back to the
$x_j$ basis, we get the operators $p_j$, given by $\langle x' | p_j | x
\rangle = (F^{\tau x_{j-1}}_{\tau x_{j+1}})_{x'1} (F^{\tau x_{j-1}}_{\tau
  x_{j+1}})_{1x}$. The total Hamiltonian is then defined as
\begin{equation} \label{eq:H-anyons}
  H_{\rm anyons} = - \epsilon \ \sum_{j=0}^{M-1} p_j \,.
\end{equation}

We are now ready to state the correspondence between the anyonic chain and our
Hamiltonian $H$~\eqref{eq:H}. The expression of the $F$-matrix is such that
the unnormalised projectors $e_j=(2\cos \frac{\pi}{5}) \, p_j$ satisfy the TL
algebra~\eqref{eq:TL-def} for $\sqrt{Q}=2\cos \frac{\pi}{5}$
\cite{anyons1}. In fact, the allowed words $x_1 \dots x_{M-1}$ are in
bijection with the $A_4$ RSOS configurations, and the operators $e_j$ act on
them as in the $A_4$ RSOS model. This exact correspondence holds in general
between the $\Zbb_k$ anyonic chain and the $A_{k+1}$ RSOS model, with the loop
weight $\sqrt{Q}=2\cos \frac{\pi}{k+2}$.

Let us recall briefly the construction of the RSOS representation of the TL
algebra~\cite{RSOS}, in the case of the $A_p$ models.  The basis states for
the vector space $\cal H$ are labelled by $M=2N$ height variables $(h_1,
\dots, h_{2N})$, such that, for all $j$:
\begin{equation}
  h_j \in \{1,\dots, p\} \,, \qquad
  |h_j-h_{j+1}| = 1 \,.
\end{equation}
Then we define the operators $(e_j)_{j=1 \dots 2N}$ acting on $\cal H$ by
their action on the basis states:
\begin{eqnarray}
  e_j|h_1 \dots h_{2N} \rangle &=& \delta_{h_{j-1},h_{j+1}}
  \sum_{|h'_j-h_{j+1}|=1} \frac{\sqrt{S(h_j) S(h'_j)}}{S(h_{j+1})}
  |h_1 \dots h'_j \dots h_{2N} \rangle \,, \label{eq:RSOS1} \\
  S(h) &=& \sin \frac{\pi h}{p+1} \label{eq:RSOS2} \,.
\end{eqnarray}
It can be shown~\cite{RSOS} that these operators satisfy the TL
algebra~\eqref{eq:TL-def} with loop weight $\sqrt{Q}=2\cos \pi/(p+1)$. In the
particular case of $\Zbb_3$ anyons, the bijection $x_j \to h_j$ reads:
\begin{eqnarray*}
  1 \to 1 \,, \ \tau \to 3 &\qquad& \text{for $j$ odd} \\
  1 \to 4 \,, \ \tau \to 2 &\qquad& \text{for $j$ even}
\end{eqnarray*}
Note that this maps the anyonic words $x_1 \dots x_{M-1}$ to the subset of
RSOS configurations such that even sites $j$ carry even heights $h_j$.
However, this subset and its complementary (where even sites carry odd
heights) are not coupled by the $e_j$. The two sectors are related
by a one-site translation. Hence the equivalence between $\Zbb_3$ anyons and
$A_4$ RSOS is valid up to a degeneracy factor of two.

\begin{figure}
  \begin{center}
    \includegraphics{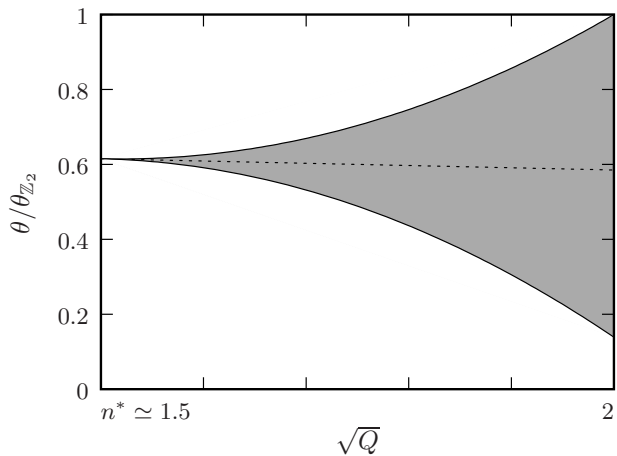}
    \caption{Phase diagram of the quadratic Temperley-Lieb
      Hamiltonian~\eqref{eq:H-TL}, using the
      parameterisation~\eqref{eq:param-K}.
      The shaded area is the gapped phase,
      governed by the MG line (dotted line). Full lines represent
      phase transitions from the MG phase to the XXZ critical phase (bottom)
      and the $\Zbb_2$ critical phase (top). The $\Zbb_2$ staggered model sits
    on the topmost horizontal line $\theta=\theta_{\Zbb_2}$.}
    \label{fig:ph-diag}
  \end{center}
\end{figure}

From this equivalence between anyonic chains and RSOS models, one understands
that $H_{\rm anyons}$~\eqref{eq:H-anyons} is just an RSOS version of the
well-studied XXZ model with anisotropy $\Delta=\sqrt{Q}/2$. It is
known~\cite{Huse} that the corresponding effective field theories are the CFT
minimal models.  More interestingly, in Ref.~\cite{anyons2}, a three-site
interaction was introduced:
\begin{equation} \label{eq:H-anyons2} H'_{\rm anyons} = \alpha
  \sum_{j=0}^{M-1} p_j + \beta \sum_{j=1}^M p'_j \,,
\end{equation}
where $p'_j$ denotes the projector onto the $1$-channel for the fusion of the
$\tau$-anyons at positions $(j,j+1,j+2)$. Hamiltonian $H'_{\rm anyons}$ is the
RSOS representation of the quadratic TL Hamiltonian~\eqref{eq:H-TL}. In the
context of anyons, only Beraha values $\sqrt{Q}=2\cos \pi/(p+1)$ with
$p=3,4,5,\ldots$ are allowed, whereas the loop formulation exists for generic
$Q$. Hence $H(K_1,K_2)$~\eqref{eq:H-TL} is really a generalisation of $H'_{\rm
  anyons}$~\eqref{eq:H-anyons2}.  Numerical studies~\cite{anyons2,IJS-TL} have
shown that the phase diagram of this model is very rich. In particular, it
contains a critical phase renormalising to the XXZ Hamiltonian, a gapped phase
of MG type (see above), and another critical phase governed by the integrable
$\Zbb_2$ model which is studied in the present paper. It was also shown in
these works that the transition between the XXZ and MG phase is in the
universality class of the dilute ${\rm O}(n)$ model, with $n=\sqrt{Q}$.
We recall the results of~\cite{IJS-TL} in Fig.~\ref{fig:ph-diag}, where we
have used the same conventions as in~\cite{anyons2} to parameterise the
interaction:
\begin{equation} \label{eq:param-K}
  \begin{array}{rcl}
    K_1 &=& 2\sqrt{Q} \ \sin \theta - \cos \theta \\
    K_2 &=& - \sin \theta \,.
  \end{array}
\end{equation}
The phase diagram is represented in the range $[\theta_{\Zbb_2},0]$, where
$\theta_{\Zbb_2}={\rm Arctan \ Q^{-1/2}} - \pi$ is the value corresponding to
the $\Zbb_2$ model.

\section{Bethe Ansatz solution}
\label{sec:bethe-ansatz}

In this section, we present the solution by Bethe Ansatz of the model
presented in Section~\ref{sec:potts}.  We find that the Bethe roots form two
{\it coupled} Fermi seas, and the elementary excitations are holes close to
the Fermi levels.  In the continuum limit, we obtain the dressed momentum and
energy~\eqref{eq:kd-ed} of the holes, and the dressed scattering
amplitudes~\eqref{eq:scatt-amp} between them. The central charge of the theory
is $c=2$. Using the Wiener-Hopf technique for the computation of finite-size
corrections (see Appendix B), we derive the conformal
spectrum~\eqref{eq:Delta}.  It has the form of a two-component Coulomb gas.

\subsection{Bethe Ansatz Equations}

We use the Algebraic Bethe Ansatz~\cite{IKB}, and we define the Bethe roots
$\alpha_j$ as: $\alpha_j = \rmi(\gamma - 2u_j)$. The Bethe Ansatz Equations
(BAE) and the eigenvalues and eigenvectors 
of the transfer matrix $t(u)$ in the $r$-particle sector are:
\begin{equation} \label{eq:BE-alpha}
  \left[ \frac{\sinh(\alpha_j - \rmi\gamma)}{\sinh(\alpha_j + \rmi\gamma)} \right]^N
  = - \rme^{2\rmi\phi} \prod_{l=1}^r 
  \frac{\sinh \half(\alpha_j-\alpha_l - 2\rmi\gamma)}
  {\sinh \half(\alpha_j-\alpha_l + 2\rmi\gamma)} \,,
\end{equation}
\begin{equation} \label{eq:Lambda}
  \Lambda(u) = \frac{1}{2^N} \left\{
  \rme^{\rmi\phi} [\sin 2(\gamma-u)]^N
  \prod_{j=1}^r
  \frac{\sinh \half \left[\rmi\gamma + (\alpha_j + 2\rmi u)\right]}
  {\sinh \half \left[\rmi\gamma - ( \alpha_j + 2\rmi u)\right]} \notag 
   + \ \rme^{-\rmi\phi} (-\sin 2u)^N
  \prod_{j=1}^r
  \frac{\sinh \half \left[\rmi\gamma - (\alpha_j + 2\rmi u - 2\rmi\gamma)\right]}
  {\sinh \half \left[\rmi\gamma + ( \alpha_j + 2\rmi u - 2\rmi\gamma)\right]}
  \right\} \,,
\end{equation}
\begin{equation} \label{eq:Psi}
  |\Psi(u_1,\dots,u_r) \rangle = B(u_1) \dots B(u_r) |0\rangle\,.
\end{equation}
In Eq.~\eqref{eq:Psi}, we have used the notations of~\cite{IKB} for the monodromy matrix elements. The Bethe states $|\Psi(u_1,\dots,u_r) \rangle$ are invariant under the two-site cyclic translation
$\rme^{-2\rmi P}$. In the very anisotropic limit $u \to 0$, the transfer matrix becomes:
\begin{equation}
  t(0) \ t\left(\frac{\pi}{2} \right) = \left(-\frac{\sin^2 2\gamma}{4}\right)^N
  \rme^{-2\rmi P}\,,
\end{equation}
and the corresponding eigenvalue is, from~\eqref{eq:Lambda}:
\begin{equation} \label{eq:k-alpha}
  \Lambda(0) \ \Lambda \left(\frac{\pi}{2} \right) = \rme^{2\rmi \phi} \left(-\frac{\sin^2 2\gamma}{4}\right)^N
  \prod_{j=1}^r \frac{\sinh(\alpha_j + \rmi\gamma)}{\sinh(\alpha_j - \rmi\gamma)} \,.
\end{equation}
The energy for the Hamiltonian~\eqref{eq:H} is the logarithmic derivative of the eigenvalue:
\begin{equation} \label{eq:E-alpha}
  E \equiv -\frac{1}{2} \sin 2\gamma 
  \ \left. \frac{{\rm d} \log[\Lambda(u)\Lambda(u+\pi/2)]}{{\rm d} u} \right|_{u=0}
  = 2N \cos 2\gamma - \sum_{j=1}^r \frac{2 \sin^2 2\gamma}{\cosh 2 \alpha_j - \cos 2\gamma} \,.
\end{equation}
Equations~\eqref{eq:k-alpha}-\eqref{eq:E-alpha} show that each Bethe root
$\alpha$ contributes to the total momentum and energy by:
\begin{equation}
  2 k_j = -\rmi \log \frac{\sinh(\alpha_j - \rmi\gamma)}{\sinh(\alpha_j + \rmi\gamma)}
  \,, \qquad
  \epsilon_j = -\frac{2 \sin^2 2\gamma}{\cosh 2 \alpha_j - \cos 2\gamma} \,.
\end{equation}

Because of the periodicity property of the Boltzmann weights $\Rc(u+\pi) = -\Rc(u)$, the Bethe states
are unchanged under $\alpha \to \alpha +2\rmi\pi$ for any of the Bethe roots. So we can restrict our 
study to the strip: $0 \leq \im \alpha < 2\pi$. The root $\alpha$ gives a negative contribution to the
energy~\eqref{eq:E-alpha} if it is of the form:
\begin{equation}
  \alpha^{(0)}_j = \lambda^{(0)}_j \quad {\rm or} \quad \alpha^{(1)}_j = \lambda^{(1)}_j + \rmi\pi \,,
\end{equation}
with $\lambda^{(a)}_j$ real. So, at low energies, the system is described by two coupled Fermi
seas $\{ \lambda^{(0)}_j \}, \{ \lambda^{(1)}_j \}$. The BAE for the $\lambda^{(a)}_j$ are:
\begin{equation} \label{eq:BAE-log}
  2N k(\lambda_j^{(a)}) = 2 \pi I_j^{(a)} + 2\phi -
  \sum_{b=0,1} \sum_{l=1}^{r^{(b)}} \Theta^{(a-b)}(\lambda_j^{(a)}-\lambda_l^{(b)}) \,,
  \quad a=0,1 \,,
\end{equation}
where $r^{(a)}$ is the number of roots $\alpha^{(a)}$.
The momentum, energy and scattering phases are given by:
\begin{equation}
  \begin{array}{lcl}
    {\displaystyle
      2k(\lambda) = -\rmi \log \frac{\sinh(\rmi\gamma-\lambda)}{\sinh(\rmi\gamma+\lambda)}
    } \,,
    &\quad& {\displaystyle 
      \Theta^{(0)}(\lambda) = -\rmi \log \frac{\sinh(\rmi\gamma+\frac{\lambda}{2})}
      {\sinh(\rmi\gamma-\frac{\lambda}{2})}
    } \,,\\
    \\
    {\displaystyle
      \epsilon(\lambda) = -\sin 2\gamma \times 2k'(\lambda) 
      = -\frac{2\sin^2 2\gamma}{\cosh 2 \lambda - \cos 2\gamma}
    } \,,
    &\quad& {\displaystyle
      \Theta^{(\pm 1)}(\lambda) = -\rmi \log \frac{\cosh(\rmi\gamma+\frac{\lambda}{2})}
      {\cosh(\rmi\gamma-\frac{\lambda}{2})}} \,.
  \end{array}
\end{equation}
The Bethe integers satisfy $I_j^{(a)} \in \half(N+r^{(a)}-1) + \mathbb{Z}$.
The total momentum and energy are:
\begin{eqnarray}
  2Q &=& \frac{2\pi}{N} \sum_{a=0,1} \sum_{j=0}^{r^{(a)}} 
  \left( I_j^{(a)} + \frac{\phi}{\pi} \right) +\pi(r^{(0)}+r^{(1)}) \,, \\
  E &=& 2N \cos 2\gamma + \sum_{a=0,1} \sum_{j=1}^{r^{(a)}} \epsilon(\lambda_j^{(a)}) \,.
\end{eqnarray}

A special case is when the Bethe roots $\{ \lambda_j^{(0)}\}, \{ \lambda_j^{(1)}\}$ are identical on the two
lines: we call these {\it symmetric states}. It is a remarkable fact that this
subset of the spectrum is exactly the complete spectrum
of the XXZ spin chain on a periodic lattice with $N$ sites:
\begin{equation} \label{eq:H-XXZ}
  H_{\rm XXZ} = -\frac{1}{2} \sum_{m=1}^N \left[
    \sigma_m^x \sigma_{m+1}^x + \sigma_m^y \sigma_{m+1}^y +\Delta_0 \ \sigma_m^z \sigma_{m+1}^z
  \right]\,,
\end{equation}
with $\Delta_0 =-\cos 2\gamma$.
Indeed, we have the identities:
\begin{equation}
  2 \kt(\lambda) = k_{\rm XXZ}(\lambda) \,,
  \quad \epsilon(\lambda) = \epsilon_{\rm XXZ}(\lambda) \,,
  \quad (\Theta^{(0)} + \Theta^{(\pm 1)})(\lambda) =\Theta_{\rm XXZ}(\lambda)\,,
\end{equation}
where quantities with the subscript `XXZ' are related to the Bethe Ansatz for XXZ.
Thus, the BAE~\eqref{eq:BAE-log} for symmetric states are equivalent to the XXZ ones,
and the energies are related by $E=2 E_{\rm XXZ}$.

\subsection{Continuum limit}

The ground state corresponds to $r^{(0)}=r^{(1)}=N/2$, with the Bethe integer distribution
(see Fig.~\ref{fig:bethe-int}a):
\begin{equation}
  I_j^{(0)} = I_j^{(1)} = -\frac{N/2+1}{2} + j \,, \quad j=1,\dots,N/2\,.
\end{equation}
The continuum limit is defined as:
\begin{equation}
  N \to \infty \,, \qquad r^{(a)}/N \to 1/2 \,.
\end{equation}
In this limit, we assume that the spacing between Bethe roots scales like $1/N$, and we
describe the Bethe root distribution by the
densities: $N \rho^{(a)}(\lambda_j^{(a)}) = 1/(\lambda_{j+1}^{(a)}-\lambda_j^{(a)})$.
We denote $[-C^{(a)},B^{(a)}]$ the interval spanned by the roots $\lambda^{(a)}_j$.
The BAE equations~\eqref{eq:BAE-log} become Lieb equations for the root densities:
\begin{equation} \label{eq:lieb}
  2 k'(\lambda) = 2\pi \rho^{(a)}(\lambda) - 
  \sum_{b=0,1} \int_{-C^{(b)}}^{B^{(b)}} d\mu \ \rho^{(b)}(\mu) \ K^{(a-b)}(\lambda-\mu)\,,
  \quad a=0,1 \,,
\end{equation}
where the kernels are given by: $K^{(a)}=(\Theta^{(a)})'$.
In the ground state, we have $C^{(0,1)},B^{(0,1)} \to \infty$, so Eq.~\eqref{eq:lieb}
can be solved by Fourier transform. The solution involves the symmetric and antisymmetric inverse
kernels $J^{(\pm)}$ (see Appendix A):
\begin{equation}
  1+\Jhat^{(\pm)}(\omega) \equiv \frac{2\pi}{2\pi -[\Khat^{(0)}(\omega) \pm \Khat^{(1)}(\omega)]} \,.
\end{equation}
The ground-state densities are $\rho^{(0)}=\rho^{(1)}=\rho_\infty$, where:
\begin{equation}
  \rho_\infty = (\delta+J^{(+)}) \star (2k')/(2\pi)
  = \frac{1}{4\gamma \cos[\pi \lambda/(2\gamma)]} \,.
\end{equation}
The symbol $\star$ denotes convolution.

An elementary excitation above the ground state consists of a hole
$\lambda_h$ in the distribution $\{\lambda_j^{(0)}\}$ or $\{\lambda_j^{(1)}\}$,
interacting with all the particles in both Fermi seas.
Let $A$ be a physical quantity defined as:
\begin{equation} \label{eq:def-A}
  A = \frac{1}{N} \sum_{a=0,1} \sum_{j=1}^{r^{(a)}} \alpha(\lambda_j^{(a)})
  \to \int_{-C^{(0)}}^{B^{(0)}} d\lambda \ \rho^{(0)}(\lambda) \alpha(\lambda)
  + \int_{-C^{(1)}}^{B^{(1)}} d\lambda \ \rho^{(1)}(\lambda) \alpha(\lambda)\,.
\end{equation}
In the presence of a hole $\lambda_h$, the variation of $A$ with respect to the
ground-state value $A_0$ is given by the dressed quantity $\alpha_d$ (see Appendix~A):
\begin{equation} \\ \label{eq:dA}
  A-A_0 = \frac{1}{N} \ \alpha_d(\lambda_h) \,,
  \qquad \alpha_d \equiv -(\delta+J^{(+)}) \star \alpha \,.
\end{equation}
The momentum and energy of a hole are thus:
\begin{equation} \label{eq:kd-ed}
  \begin{array}{rcl}
    2k_d(\lambda) &=& -2 {\rm Atan} \left[ 
      \tanh \left( \frac{\pi \lambda}{4\gamma} \right)
    \right] \,, \\
    \\
    \epsilon_d(\lambda) &=& \displaystyle{
      \frac{\pi \sin 2\gamma}{2\gamma \cosh[\pi \lambda/(2\gamma)]}
      } \,.
  \end{array}
\end{equation}
In the region $\lambda \to \infty$, the dressed momentum is close to the value $-\pi/2$, and
the dispersion relation is linear, with Fermi velocity $v$:
\begin{equation}
  \epsilon_d \simeq -v \ \left(2k_d+ \halfpi \right) \,,
  \qquad v=\frac{\pi \sin 2\gamma}{2\gamma} \,.
\end{equation}
Hence, hole excitations are gapless, and the theory is critical.

\subsection{Dressed scattering amplitudes}
\label{sec:scat-holes}

In the presence of holes, the root densities $\rho^{(a)}$ coexist with the densities of holes
$\rho^{(a)}_h$.
The Lieb equations~\eqref{eq:lieb} become:
\begin{equation}
  \left\{ \begin{array}{rcl}
    2\pi (\rho^{(0)}+\rho_h^{(0)}) &=& 2k' + K^{(0)} \star \rho^{(0)} + K^{(-1)} \star \rho^{(1)} \\
    2\pi (\rho^{(1)}+\rho_h^{(1)}) &=& 2k' + K^{(1)} \star \rho^{(0)} + K^{(0)} \star \rho^{(1)} \,.
  \end{array} \right.
\end{equation}
These coupled equations can be rewritten in terms of $\rho^{(a)}+\rho^{(a)}_h$ and $\rho^{(a)}_h$:
\begin{equation}
  \left\{ \begin{array}{rcl}
    \rho^{(0)}+\rho^{(0)}_h &=& \rho_\infty - J^{(0)} \star \rho^{(0)}_h - J^{(-1)} \star \rho^{(1)}_h \\
    \rho^{(1)}+\rho^{(1)}_h &=& \rho_\infty - J^{(1)} \star \rho^{(0)}_h - J^{(0)} \star \rho^{(1)}_h \,,
  \end{array} \right.
\end{equation}
where the kernels $J^{(0)},J^{(\pm 1)}$ are defined as:
\begin{equation}
  J^{(0)} \equiv \half(J^{(+)}+J^{(-)}) \,, \qquad J^{(\pm 1)} \equiv \half(J^{(+)}-J^{(-)}) \,.
\end{equation}
The Fourier transforms of the kernels $J^{(0)},J^{(\pm 1)}$ are:
\begin{equation} \label{eq:scatt-amp}
  \begin{array}{rcl}
    \Jhat^{(0)}(\omega) &=& \displaystyle{
      \frac{\sinh(\pi-3\gamma)\omega}{2\cosh \gamma \omega \sinh(\pi-2\gamma)\omega}
      } \,, \\
      \\
    \Jhat^{(\pm 1)}(\omega) &=& \displaystyle{
      -\frac{\sinh \gamma\omega}{2\cosh \gamma \omega \sinh(\pi-2\gamma)\omega}
      } \,.
  \end{array}
\end{equation}

\subsection{Central charge and conformal dimensions}

\begin{figure}
  \begin{center}
    \includegraphics[scale=0.8]{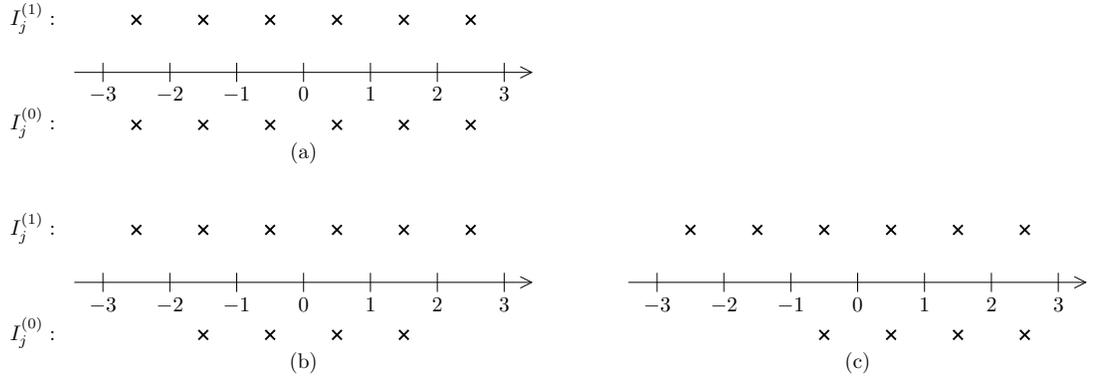}
    \caption{Example of Bethe integer distributions for $N=12$. (a) Ground state; 
      (b) Magnetic excitation $m^{(0)}=2$; 
      (c) Combined magnetic-electric excitation $m^{(0)}=2, e^{(0)}=1$.}
    \label{fig:bethe-int}
  \end{center}
\end{figure}

The low-energy spectrum consists of `electromagnetic' excitations above the ground-state distribution,
similarly to the XXZ case~\cite{Alcaraz-etal}. In the present case, the Bethe integers distibutions
$\{ I_j^{(0)}\}, \{ I_j^{(1)}\}$ can be chosen independently, as depicted in
Fig.~\ref{fig:bethe-int}b-\ref{fig:bethe-int}c.
A magnetic excitation
consists in removing $m^{(0)}$ (resp. $m^{(1)}$) roots of type $\alpha_j^{(0)}$
(resp. $\alpha_j^{(1)}$) from the ground state, while keeping the Bethe integer
distrubutions $\{ I_j^{(0,1)} \}$ symmetric around zero.
An electric excitation consists in shifting all integers $I_j^{(0)}$ (resp. $I_j^{(1)}$)
by $e^{(0)}$ (resp. $e^{(1)}$).

The central charge and critical exponents are obtained from finite-size corrections to 
the total energy and momentum:
\begin{eqnarray}
  E_0 &\simeq& N e_\infty - v \times \frac{\pi c}{6N}\,, \label{eq:E-gs} \\
  E - E_0 &\simeq& v \times \frac{2\pi (\Delta+\Deltab)}{N}\,, \\
  Q &=& \frac{2\pi(\Delta-\Deltab)}{N}\,,
\end{eqnarray}
where $E_0$ is the ground-state energy.

We start by discussing the untwisted case $\phi=0$.
The ground state is a symmetric state, and thus the ground-state energy is twice
that of the XXZ spin-chain~\eqref{eq:H-XXZ}. The latter has a Fermi velocity $v$ and central charge one. Using
Eq.~\eqref{eq:E-gs}, the central charge of the staggered model is then: $c=2$.
The finite-size corrections to the energies for the electromagnetic excitations
are computed from the Bethe-Ansatz solution in Appendix~B.
They yield the conformal dimensions:
\begin{equation} \label{eq:Delta}
  \begin{array}{rcl}
    \Delta_{em,\et \mt} &=& \displaystyle{
      \frac{1}{8} \left( \frac{e}{\sqrt{2g}} + m \sqrt{2g} \right)^2 
      + \frac{1}{8} (\et+\mt)^2
    } \,,  \\
    \\
    \Deltab_{em,\et \mt} &=& \displaystyle{
      \frac{1}{8} \left( \frac{e}{\sqrt{2g}} - m \sqrt{2g} \right)^2
      + \frac{1}{8} (\et-\mt)^2
    } \,, \\
    \\
    g &=& \displaystyle{ \frac{\pi-2\gamma}{2\pi} } \,, \quad 0 < g < \half \,,
  \end{array}
\end{equation}
where:
\begin{equation}
  \begin{array}{lcl}
    e = e^{(0)}+e^{(1)} &\quad& m = m^{(0)}+m^{(1)} \\
    \\
    \et = e^{(0)}-e^{(1)} &\quad& \mt = m^{(0)}-m^{(1)}\,.
  \end{array}
\end{equation}

When the twist $\phi$ is not zero, the above exponents are still correct, with the change $e \to e +2 \phi/\pi$. In particular, the staggered Potts model corresponds to a twist $\phi = \gamma = \pi e_0$. The ground state has an electric charge $e=2e_0$, with exponents $\Delta=\Deltab = e_0^2/(4g)$, so the effective central charge is:
\begin{equation} \label{eq:cc}
  c_{\rm tw} = 2 - \frac{6 e_0^2}{g} \,, \qquad e_0 = \frac{\gamma}{\pi} = \half-g \,.
\end{equation}

\subsection{Application to the calculation of critical exponents}

\begin{figure}
  \begin{center}
    \includegraphics[scale=1]{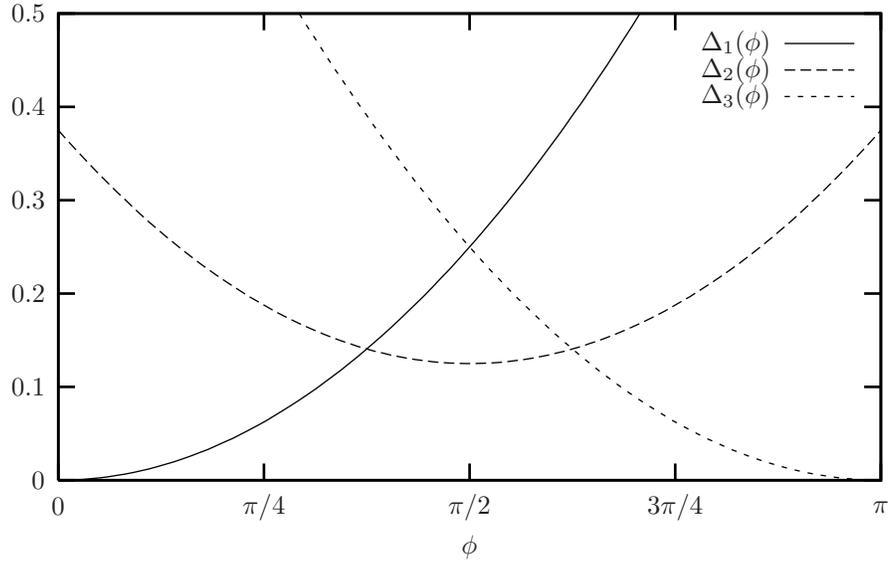}
    \caption{Lowest conformal dimensions in the twisted sector, for $g=1/4$.}
    \label{fig:Delta123}
  \end{center}
\end{figure}

In the loop formulation, we obtain the $k$-leg dimensions as follows. For any system size $N$, the number of legs $k$ must be even, and the conformal dimensions are defined with respect to the twisted ground state:
\begin{equation}
  h_k = \bar{h}_k = \Delta_k - \frac{e_0^2}{4g} \,.
\end{equation}
The $k$-leg dimension $\Delta_k$ corresponds to a magnetic defect $m=k/2$, 
with a minimal value for $\mt$ and electric charges $e=\et=0$ (no background charge).
There are then two distinct cases:
\begin{equation} \label{eq:hk}
  h_k = \begin{cases}
    {\displaystyle \frac{g k^2}{16} - \frac{e_0^2}{4g}} & \text{if $k \equiv 0 \ [4]$} \\
    {\displaystyle \frac{g k^2}{16} + \frac{1}{8} - \frac{e_0^2}{4g}} & \text{if $k \equiv 2 \ [4]$} \,.
  \end{cases}
\end{equation}
Similarly, the magnetic exponent of the staggered Potts model is defined with respect to the twisted
ground state:
\begin{equation}
  h_H = \bar{h}_H = \Delta_H - \frac{e_0^2}{4g} \,.
\end{equation}
The magnetic dimension $\Delta_H$ corresponds to a twist $\phi=\pi/2$, which
forbids any non-contractible loop around the cylinder. Before we obtain $\Delta_H$, we need
to discuss the conformal dimension for the sector $m=\mt=0$ with a general twist $\phi$.
In the regime $0<\phi<\pi$, the lowest dimensions are:
\begin{eqnarray}
  \Delta_1(\phi) &=& \Delta_{(\phi/\pi,0),(0,0)} = \frac{(\phi/\pi)^2}{4g} \,, \\
  \Delta_2(\phi) &=& \Delta_{(\phi/\pi-1,0),(1,0)} = \frac{(\phi/\pi-1/2)^2}{4g} + \frac{1}{8} \,, \\
  \Delta_3(\phi) &=& \Delta_{(\phi/\pi-2,0),(0,0)} = \frac{(1-\phi/\pi)^2}{4g} \,.
\end{eqnarray}
The lowest dimension is respectively $\Delta_1,\Delta_2,\Delta_3$ on the intervals 
$[0,\phi_0], \ [\phi_0, 1-\phi_0], \ [1-\phi_0,\pi]$, where $\phi_0 = \pi(1+2g)/4$. See Fig.~\ref{fig:Delta123}. In particular, for $\phi=\pi/2$, we get $\Delta_H=\Delta_2(\pi/2)=1/8$, and thus:
\begin{equation}
  h_H = \frac{1}{8} - \frac{e_0^2}{4g} \,.
\end{equation}

\subsection{Numerical checks}

We have verified the above expressions for the effective central charge
$c_{\rm tw}$ and the $k$-leg exponents $h_k$ by numerical diagonalisation
of the transfer matrix at the pseudo-isotropic point $u=\frac{\gamma}{2}$.

As usual, the critical exponents can be extracted from the finite-size
corrections in $N$ to the dominant eigenvalues in the various sectors
labelled by $k$. We consider the geometry of a strip of width $2N$
strands with periodic boundary conditions in the transverse direction.
Estimates for $c_{\rm tw}$ (resp.\ $h_k$) are then obtained from fits
involving three (resp.\ two) different sizes $N$.  We use even $N$
throughout. Odd $N$ introduces a twist that leads to different effective
exponents that we do not consider any further.

\begin{figure}
  \begin{center}
    \includegraphics[scale=0.3]{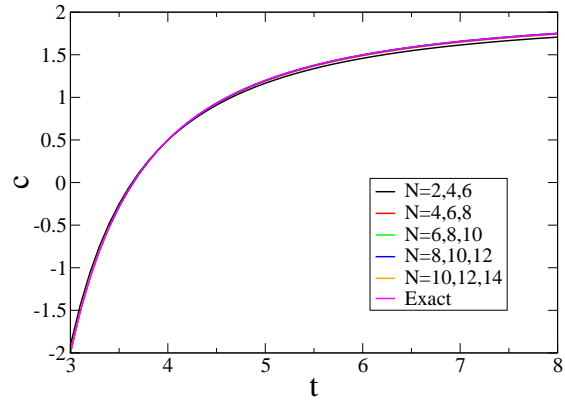}
    \caption{Numerical estimates of the effective central charge $c_{\rm tw}$,
             as compared with the exact expression (\ref{eq:cc}).}
    \label{fig:cc}
  \end{center}
\end{figure}

\begin{figure}
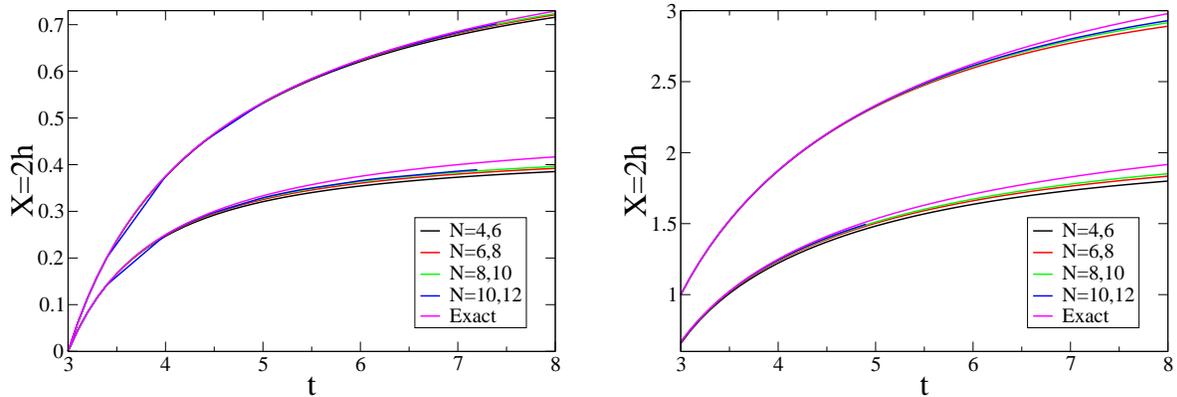

  \begin{center}
    \includegraphics[scale=0.3]{s2_4.eps} \qquad
    \includegraphics[scale=0.3]{s6_8.eps}
    \caption{Numerical estimates of the $k$-leg exponents $X=2h_k$ with
             $k=2,4,6,8$,
             as compared with the exact expression (\ref{eq:hk}).}
    \label{fig:hk}
  \end{center}
\end{figure}

It is convenient to parameterise $\gamma = \frac{\pi}{t}$ through a
new parameter $t$.  The results for $c_{\rm tw}$ with $N \le 14$ are shown in
Fig.~\ref{fig:cc}. The agreement with (\ref{eq:cc}) is
excellent. Results for $X_k = 2h_k$ with $N \le 12$ are given in
Fig.~\ref{fig:hk} for $k=2,4,6,8$. The agreement with (\ref{eq:hk}) is very
satisfactory, especially when $k \equiv 0 \ [4]$.

\section{Toroidal partition functions}
\label{sec:Z-torus}

In this section, we use the results from Section~\ref{sec:bethe-ansatz} to construct
explicitly the continuum partition function $Z$ of the statistical model on a torus.
Assuming that the conformal spectrum~\eqref{eq:Delta} obtained from the analysis of the BAE
is complete, we sum the conformal characters over all possible conformal dimensions to obtain $Z$.
The resulting expression~\eqref{eq:Zg-Ising} for $Z$ shows that the continuum limit of the model
consists in one boson and two Majorana fermions, which decouple in the bulk and
couple only through boundary conditions. We discuss only the untwisted case
here, leaving the twisted and Potts model cases (including the study of
particular values  of $Q$) to Appendix~D.

We denote by $\tau$ the modular ratio of the torus, and we write $q=\rme^{2\rmi \pi \tau}$.
Other notations are defined in Appendix~C.
The primary states of the corresponding CFT have conformal weights
$\Delta_{e m,\et \mt}$ and $\Deltab_{e m,\et \mt}$ given by the Bethe
Ansatz results \eqref{eq:Delta}, where the charges satisfy the parity
conditions:
\begin{equation} \label{eq:parity}
  e + \et \in 2\Zbb \,, \qquad m + \mt \in 2\Zbb \,.
\end{equation}
The partition function on the torus is given by the sum of the generic conformal characters $\chi_\Delta,\bar{\chi}_{\Deltab}$:
\begin{equation}
  Z(g) = {\rm Tr} \left( q^{L_0-c/24} \ \qb ^{\bar{L}_0-c/24} \right)
  = \sum_{\Delta,\Deltab} 
  \chi_{\Delta}(q) \bar{\chi}_{\Deltab}(\qb) \,,
\end{equation}
where the sum is over all possible primary states, and $\chi_\Delta$ is the trace of $q^{L_0-c/24}$ over the descendants of the primary state $\Phi_\Delta$:
\begin{equation}
  \chi_\Delta(q) = {\rm Tr}_\Delta \ q^{L_0-c/24} \,.
\end{equation}
The character $\chi_\Delta$ can be inferred from the possible Bethe integer distributions. Starting from an electromagnetic excitation with dimension $\Delta$, we can create vacancies, by shifting the largest Bethe integer $I_j \to I_j+n, n \geq 0$. This vacancy state has dimension $\Delta+n$. These vacancies can be combined, and the state with shifts $(n_1,\dots,n_k)$ has dimension $\Delta+n_1+\dots+n_k$. Furthermore, vacancies can be introduced {\it independently} on the two lines $I_j^{(0)}, I_j^{(1)}$. Let us denote $p(n)$ the number of partitions of the integer $n$. We have:
\begin{equation} \label{eq:chi}
  \chi_\Delta(q) = \sum_{n^{(0)},n^{(1)} \geq 0} p(n^{(0)}) p(n^{(1)}) q^{\Delta+n^{(0)}+n^{(1)}-c/24}
  = \frac{q^{\Delta+(2-c)/24}}{\eta(\tau)^2} \,,
\end{equation}
where $\eta(\tau)$ is the Dedekind function~\eqref{eq:eta}.
Using \eqref{eq:chi} with $c=2$ and the parity conditions~\eqref{eq:parity}, we obtain:
\begin{equation}
  Z(g) = \frac{1}{|\eta(\tau)|^4} \left(
    \sum_{\myscript{m,\mt \ {\rm even}}{e,\et \ {\rm even}}}
    + \sum_{\myscript{m,\mt \ {\rm even}}{e,\et \ {\rm odd}}}
    + \sum_{\myscript{m,\mt \ {\rm odd}}{e,\et \ {\rm even}}}
    + \sum_{\myscript{m,\mt \ {\rm odd}}{e,\et \ {\rm odd}}}
  \right) q^{\Delta_{em,\et\mt}} \qb^{\Deltab_{em,\et\mt}} \,.
\end{equation}
Using the Poisson summation~\eqref{eq:Poisson}, this can be written:
\begin{equation} \label{eq:Zg}
  Z(g) = 2 \left(
    A \sum_{m,m' \ {\rm even}}
    + B \sum_{m \ {\rm even}, m' \ {\rm odd}}
    + C \sum_{m \ {\rm odd}, m' \ {\rm even}}
    + D \sum_{m,m' \ {\rm odd}}
  \right) Z_{m,m'}(g) \,,
\end{equation}
where:
\begin{equation}
  \begin{array}{rclrcl}
    A &=& {\displaystyle \sum_{m,m' \, {\rm even}} Z_{m,m'}(1/2)} \,, & \qquad
    B &=& {\displaystyle \sum_{m \, {\rm even}, m' \, {\rm odd}} Z_{m,m'}(1/2)} \,, \\
    \\
    C &=& {\displaystyle \sum_{m \, {\rm odd}, m' \, {\rm even}} Z_{m,m'}(1/2)} \,, & \qquad
    D &=& {\displaystyle \sum_{m,m' \, {\rm odd}} Z_{m,m'}(1/2)} \,,
  \end{array}
\end{equation}
and $Z_{m,m'}(g)$ is the bosonic partition function with defects $m,m'$
(see~\eqref{eq:Zmm}).

The partition sums $A,B,C,D$ can, in turn, be expressed in terms of the Jacobi
ones~\eqref{eq:Znu}, using~\eqref{eq:Poisson} again:
\begin{equation} \label{eq:ABCD}
  \begin{array}{rclrcl}
  A &=& \frac{1}{4} ( Z_2^2 +Z_3^2 + Z_4^2 ) \,, & \qquad
  B &=& \frac{1}{4} ( -Z_2^2 +Z_3^2 + Z_4^2 ) \,, \\
  \\
  C &=& \frac{1}{4} ( Z_2^2 +Z_3^2 - Z_4^2 ) \,, & \qquad
  D &=& \frac{1}{4} ( Z_2^2 -Z_3^2 + Z_4^2 ) \,.
  \end{array}
\end{equation}
Using the transformation of Jacobi and Coulombic partition functions under
modular transformations, one can
show easily that the expression~\eqref{eq:Zg} is modular invariant.
Let $\Zc(r,r')$ be the partition function of the Ising model on a torus with
respective boundary conditions on the spins $\sigma$ in the two directions of
the torus:
\begin{equation}
  \sigma \to (-1)^r \sigma \,, \qquad \sigma \to (-1)^{r'} \sigma
  \,, \qquad r,r' \in \{0,1\}^2 \,.
\end{equation}
Using the relation~\eqref{eq:Z-Zc} between $\Zc(r,r')$ and $Z_\nu$, the partition sums 
$A,B,C,D$ are written in terms of the $\Zc(r,r')$:
\begin{eqnarray}
  A &=& \frac{1}{4} \left[ \Zc(0,0)^2 + \Zc(0,1)^2 + \Zc(1,0)^2 + \Zc(1,1)^2 \right] \,, 
  \label{eq:A-Ising} \\
  B &=& \frac{1}{2} \left[ \Zc(0,0) \Zc(0,1) - \Zc(1,0) \Zc(1,1) \right] \,, \\
  C &=& \frac{1}{2} \left[ \Zc(0,0) \Zc(1,0) - \Zc(0,1) \Zc(1,1) \right] \,, \\
  D &=& \frac{1}{2} \left[ \Zc(0,0) \Zc(1,1) - \Zc(0,1) \Zc(1,0) \right] \,.
  \label{eq:D-Ising}
\end{eqnarray}
Hence, from \eqref{eq:Zg} and (\ref{eq:A-Ising}--\ref{eq:D-Ising}), the partition function
$Z(g)$ reads:
\begin{equation} \label{eq:Zg-Ising}
  Z(g) = \frac{1}{2} \sum_{\myscript{r_1,r_2}{r'_1,r'_2}}
  (-1)^{r_1 r'_2 + r'_1 r_2} \ \Zc(r_1,r'_1) \ \Zc(r_2,r'_2)
  \sum_{\myscript{m \equiv r_1+r_2 \ [2]}{m' \equiv r'_1+r'_2 \ [2]}}
  Z_{m,m'}(g) \,.
\end{equation}
The degrees of freedom contained in $Z(g)$ are a compact boson 
$\varphi$ (see~\eqref{eq:A-boson}) with 
coupling constant $g=(\pi-2\gamma)/(2\pi)$, and two sets of Ising spins $\sigma_1, \sigma_2$.
The boundary defects for $\varphi, \sigma_1, \sigma_2$ are respectively $(m,m'), (r_1,r'_1), (r_2,r'_2)$, and obey parity conditions, as shown in~\eqref{eq:Zg-Ising}.
Apart from these conditions, the three degrees of freedom $\varphi, \sigma_1, \sigma_2$
are decoupled. These results are very similar to what was found in~\cite{SCFT} for
a lattice model related to $N=1$ superconformal theories, where only one Ising
spin was present.

Like it was done in~\cite{SCFT} for the 19-vertex model, here we can also
identify the degrees of freedom $\varphi, \sigma_1, \sigma_2$ in the lattice
model. For this purpose, we consider the vertex model defined by
the block $\RRc$-matrix (see Fig.~\ref{fig:block-Rmat}). It was shown
in~\cite{IJS-PAF} that there are 38 possible vertices. Each edge can be in one
of four states: $\uparrow, \downarrow, |, \|$. Let $N_\alpha({\bf r})$ be the
number of edges adjacent to the site $\bf r$, which are in the state $\alpha$.
An essential property of the model, arising from the combination of the
magnetisation conservation and $\Zbb/2\Zbb$ symmetry, is that $N_|$ and
$N_{\|}$ are both even for every vertex. Thus, for a given lattice
configuration, the lines formed by the $|$ and $\|$ edges can be viewed as the
domain walls of two distinct Ising models, both living on the dual lattice.
The remaining edges carry arrows, which define a height (SOS) model on the
dual lattice. Although these three degrees of freedom are coupled in the lattice
model, our results on the continuum partition function show that they decouple
in the continuum limit, except for their boundary conditions, which keep track
of the parity of domain walls and arrows around each direction of the torus.

\section{Integrable massive deformation}
\label{sec:massive}

In this Section, we follow the approach of~\cite{RS} to construct a massive
deformation of the lattice model, and study its excitation spectrum. Using the
dressed scattering amplitudes, we obtain partly the $S$-matrix for elementary
excitations. Then we use known results from~\cite{FSZ} on this $S$-matrix
to conjecture a TBA diagram, and we use the TBA equations to calculate the
ground-state energy scaling function. In the UV limit, we retrieve the results
from Section~\ref{sec:bethe-ansatz}. Finally, we use these results to propose an
effective QFT for the massive deformation, which is a complex version of the
$C_2^{(2)}$ Toda theory.

\subsection{Massive integrable deformation on the lattice}

We now consider a deformation of our model where the spectral parameters
acquire an extra staggering, this time in the imaginary direction. We choose
the pattern
$u+\rmi\Lambda/2,u-\rmi\Lambda/2,u+\pi/2+\rmi\Lambda/2,u+\pi/2-\rmi\Lambda/2$. This
kind of construction has been widely used to induce an integrable massive
deformation from integrable lattice models \cite{RS,DV}. We obtain a modified
set of Bethe equations:
\begin{equation}
  \left\{ \begin{array}{l}
    2\pi (\rho+\rho_{h})^{(0)} (\lambda) =
    2k'(\lambda+\Lambda) + 2k'(\lambda-\Lambda)
    + (K^{(0)} \star \rho^{(0)})(\lambda) + (K^{(-1)} \star \rho^{(1)})(\lambda) \\
    \\
    2\pi (\rho+\rho_{h})^{(1)} (\lambda) =
    2k'(\lambda+\Lambda) + 2k'(\lambda-\Lambda)
    + (K^{(1)} \star \rho^{(0)})(\lambda) + (K^{(0)} \star \rho^{(1)})(\lambda) \,.
  \end{array} \right.
\end{equation}
To explore the corresponding physics, we write what is often called the
physical equations, that is, the equations describing scattering of dressed
excitations. Since the ground state is obtained by filling up the $\rho^{(0)}$
and $\rho^{(1)}$ lines, this is easily done by reexpressing the equations so
that the densities of holes appear on the right-hand side (see
Section~\ref{sec:scat-holes}). We find:
\begin{equation} \label{eq:BAE-phys}
  \left\{ \begin{array}{l}
    2\pi (\rho+\rho_{h})^{(0)} = s + \Phi^{(0,0)} \star \rho^{(0)}_{h}
    + \Phi^{(0,1)}\star\rho^{(1)}_{h} \\
    \\
    2\pi (\rho+\rho_{h})^{(1)} = s + \Phi^{(1,0)}\star\rho^{(0)}_{h}
    + \Phi^{(1,1)}\star\rho^{(1)}_{h} \,, 
  \end{array} \right.
\end{equation}
where $\Phi^{(a,b)}=-2\pi J^{(a-b)}$, and:
\begin{equation}
  s(\lambda)=\frac{\pi}{2\gamma} \left[
    \frac{1}{\cosh \frac{\pi}{2\gamma} (\lambda-\Lambda)}
    +\frac{1}{\cosh \frac{\pi}{2\gamma} (\lambda+\Lambda)}
  \right] \,.
\end{equation}
This function has tails at $|\lambda| \gg \Lambda$ where $s(\lambda)$ decays
exponentially as in the massless case. These describe a `ghost' of the initial
massless theory, whose physics does not depend on $\Lambda$, and which we will
not discuss in the following. It decouples entirely from the region where
$|\lambda| \ll \Lambda$, which is of interest to us. In this region, we have

\begin{equation}
  s(\lambda) \approx \frac{2\pi}{\gamma} \rme^{-\pi\Lambda/(2\gamma)}\cosh \frac{\pi\lambda}{2\gamma} \,,
\end{equation}
so the corresponding momentum and energy are
\begin{eqnarray}
  2k_d(\lambda) &=& -\int_0^\lambda s(\mu) d\mu
  \approx -4 \rme^{-\pi\Lambda/(2\gamma)} \sinh \frac{\pi\lambda}{2\gamma} \,, \\
  \epsilon_d(\lambda) &=& \sin 2\gamma \ s(\lambda)
  \approx 4v \ \rme^{-\pi\Lambda/(2\gamma)} \cosh \frac{\pi\lambda}{2\gamma} \,.
\end{eqnarray}
We thus obtain a massive relativistic spectrum, as happens systematically in this kind of construction. 
The mass is given by:
\begin{equation} \label{eq:mass}
  \mu = 4 \exp \left( -\frac{\pi \Lambda}{2\gamma} \right) \,.
\end{equation}
The question is then, what kind of scattering theory do we obtain, and what quantum field theory does it correspond to?

\subsection{Scattering theory}

To answer the above question, we start by reinterpreting the kernels $\Phi^{(a,b)}$ as derivatives of
scattering phases between basic particles. We will now denote the holes in the $0,1$ sea by the labels 
$0-$ and $1-$. If we rescale the parameters $\lambda, \omega$ to:
\begin{equation}
  \theta = \frac{\pi}{2\gamma} \times \lambda \,,
  \qquad k = \frac{2\gamma}{\pi} \times \omega \,,
\end{equation}
and set $t = \pi/\gamma$, we obtain, up to a constant phase which will be obtained below:
\begin{equation}
  \begin{array}{l}
    \displaystyle{
      S^{(0,0)}_{--}(\theta)= S^{(1,1)}_{--}(\theta) \propto
      \exp \left[
        \frac{\rmi}{2} \int_{-\infty}^\infty \frac{dk}{k}
        \sin k\theta \frac {\sinh \frac{k\pi}{2}(t-3)}
        {\sinh \frac{k\pi}{2}(t-2) \cosh \frac{k\pi}{2}}
      \right] \,,
    } \\
    \\
    \displaystyle{
      S_{--}^{(0,1)}(\theta) = S_{--}^{(1,0)}(\theta) \propto 
      \rmi \ \exp \left[
        -\frac{\rmi}{2} \int_{-\infty}^\infty \frac{dk}{k}
        \sin k\theta \frac{\sinh \frac{k\pi}{2}}
        {\sinh \frac{k\pi}{2}(t-2) \cosh \frac{k\pi}{2}}
      \right] \,.
    }
  \end{array} 
\end{equation}
These two $S$-matrix elements can be interpreted in terms of the scattering matrix
$\Sc_{ij}(\beta_{\rm SG};\theta)$ of the Sine-Gordon (SG) model~\cite{ZZ}, with action:
\begin{equation}
  A_{\rm SG}[\varphi] = \int \left[ \half \partial_\nu \varphi \partial_\nu \varphi 
    + \frac{\mu_0^2}{\beta_{\rm SG}^2} \cos(\beta_{\rm SG} \varphi) \right] \,.
\end{equation}
If we set:
\begin{equation} \label{eq:param-scat}
  \frac{\beta_{\rm SG}^2}{8\pi}=\frac{t-2}{t-1} \,,
\end{equation}
then $S^{(0,0)}_{--}$ is the kink-kink (or antikink-antikink) scattering element for the 
SG model~\cite{ZZ}:
\begin{equation}
  S^{(0,0)}_{--}(\theta) = S^{(1,1)}_{--}(\theta) = \Sc_{--}(\beta_{\rm SG};\theta) \,.
\end{equation}

It is natural to identify the holes as two types of antikinks ($0,1$). We expect the scattering theory to 
contain also a corresponding doublet of kinks, with a full scattering within the $(0,0)$ and $(1,1)$ sectors 
described by two copies of the Sine-Gordon $S$-matrix. 

We now observe that the kernels $S^{(0,1)}_{--}, S^{(1,0)}_{--}$ are related
to the SG scattering matrix $\til{\Sc}$ with an imaginary shift in the
rapidity $\theta$~\cite{FSZ}. The scattering theory defined by
$S^{(0,0)}=S^{(1,1)}=\Sc,S^{(0,1)}=S^{(1,0)}=\til{\Sc}$ was introduced
in Ref.~\cite{FSZ}, where it was proposed as the scattering theory for left/right
(L/R) massless particles describing the flow between minimal models of CFT
under a perturbation by the $\Phi_{13}$ primary operator. In Ref.~\cite{FSZ}, using
the unitarity and crossing conditions, the normalising factors for the
$S$-matrix were computed. The resulting scattering theory is:
\begin{eqnarray}
  \hbox{Four basic particles} &:& 0+, 0-, 1+, 1- \nonumber \\
  S^{(0,0)}(\theta) = S^{(1,1)}(\theta) &=& \Sc(\beta_{\rm SG};\theta) \nonumber \\
  S^{(0,1)}(\theta) &=& \frac{\til{Z}(\theta)}
  {Z \left( \theta + \rmi\pi \frac{t-2}{2} \right)}
  \ \Sc \left(\beta_{\rm SG};\theta + \rmi\pi \frac{t-2}{2} \right) \nonumber\\
  S^{(1,0)}(\theta) &=& \frac{-\til{Z}(\theta)}
  {Z \left( \theta - \rmi\pi \frac{t-2}{2} \right)}
  \ \Sc \left(\beta_{\rm SG};\theta - \rmi\pi \frac{t-2}{2} \right) \,,
  \label{eq:Smain}
\end{eqnarray}
where the normalisation factors read:
\begin{equation}
  \begin{array}{l}
    \displaystyle{
      Z(\theta) = \frac{1}{\sinh \frac{\rmi \pi-\theta}{t-2}}
      \exp \left[ \frac{\rmi}{2} \int_{-\infty}^\infty \frac{dk}{k}
        \sin k\theta \frac{\sinh \frac{k\pi}{2}(t-3)}
        {\sinh \frac{k\pi}{2}(t-2) \cosh \frac{k\pi}{2}}
      \right] \,,
    } \\
    \\
    \displaystyle{
      \til{Z}(\theta) = \frac{1}{\cosh \frac{\rmi \pi-\theta}{t-2}}
      \exp \left[ -\frac{\rmi}{2}\int_{-\infty}^\infty \frac{dk}{k}
        \sin k\theta \frac{\sinh \frac{k\pi}{2}}
        {\sinh \frac{k\pi}{2}(t-2) \cosh \frac{k\pi}{2}}
      \right] \,.
    }
  \end{array}
\end{equation}
From~\eqref{eq:param-scat}, we see that the SG $S$-matrices are in the attractive regime for $t\in [2,3]$ and repulsive regime otherwise. We stress that $0,1$ are not antiparticles of each other. 

\subsection{Ground-state energy}

The scaling function for the ground-state energy is the relevant object to describe the RG flow
of a scattering theory. We consider the system on a finite circle of circumference $R$. Then
the ground-state energy $E(\mu,R)$ has the scaling form:
\begin{equation}
  E(\mu,R) = \frac{2\pi}{R} F(\mu R) \,,
\end{equation}
where $\mu$ is the mass of the elementary particles, given in~\eqref{eq:mass}.

In the present case, the ground-state energy can be obtained simply,
using the following identity on the dressed kernels:
\begin{equation} \label{eq:id-Phi}
  \hat{\Phi}^{(0,0)}(\omega) + \hat{\Phi}^{(0,1)}(\omega)
  = \frac{2\pi \sinh(\pi-4\gamma)\omega/2}
  {2\cosh\gamma\omega\sinh(\pi-2\gamma)\omega/2} \,,
\end{equation}
The right-hand side of~\eqref{eq:id-Phi} is exactly, in terms of the same rapidity $\theta$, the Sine-Gordon kernel but for yet another value of the coupling, given by
\begin{equation}
  \frac{\til{\beta}_{\rm SG}^2}{8\pi} =
  \frac{t-2}{t} \,.
\end{equation}
In other words, 
\begin{equation}
  S^{00}_{--}(\theta) \ S^{01}_{--}(\theta) = 
  \Sc_{--}(\til{\beta}_{\rm SG}; \theta) \,.
\end{equation}
Assuming that the symmetry is not broken between the two types of roots in the ground state, it
follows immediately that the ground-state energy (calculated, {\it e.g.}, by the method of Ref.~\cite{DV})
is twice the ground-state energy of the Sine-Gordon model with the same mass for the kinks,
and at this renormalised value of the coupling:
\begin{equation} \label{eq:GS-energies}
  E(\mu,R) = 2 \Ec(\til{\beta}_{\rm SG}; \mu,R) \,.
\end{equation}
This result is in fact quite obvious if we recall that symmetric solutions to the Bethe equations satisfy precisely the same system as in the XXZ chain, whose staggering produces the Sine-Gordon theory in the continuum limit.\footnote{The careful reader might worry about the role of $\Lambda$ in both points of view. The staggering in the equivalent XXZ system involves $2\Lambda$, but the anisotropy is also doubled, so the physical mass remains the same.}
 
\begin{figure}
  \begin{center}
    \includegraphics{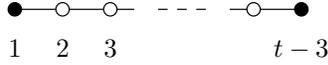}
    \caption{The TBA diagram for the RSOS restrictions of our theory.
      The leftmost and rightmost nodes are both massive.}
    \label{fig:TBA-RSOS}
  \end{center}
\end{figure}

Of course, it would be more satisfactory to establish the result
\eqref{eq:GS-energies} directly from the scattering theory. We obtain this in
the RSOS version of the model (for $t$ integer). The ground-state energy
$E(\mu,R)$ is generally obtained by the Thermodynamic Bethe Ansatz (TBA) for
relativistic scattering theories, introduced in Ref.~\cite{TBA-RFST}. 
The idea of the method is to consider a Euclidean theory on a semi-infinite
cylinder of dimensions $R \times L$, and to write the partition function in two ways:
\begin{equation}
  Z(R,L) = \exp \left[-E(\mu,R) L \right] = {\rm Tr} \left(\rme^{-R H_L}\right) \,,
\end{equation}
where $H_L$ is the Hamiltonian on an infinite domain when $L \to \infty$. The problem of computing
$E(\mu,R)$ is thus equivalent to the computation of the free energy on an infinite domain, at finite
temperature $1/R$. So one has to find the density of elementary particles which satisfies the
BAE~\eqref{eq:BAE-phys}, and maximises the free energy at temperature~$1/R$. This results in
the non-linear integral equations and the ground-state energy, given in terms
of the pseudo-energies $\epsilon_a$~\cite{TBA-RFST}:
\begin{equation} \label{eq:TBA}
  \begin{array}{c}
    \displaystyle{\epsilon_a(\theta) = \mu_a R \cosh \theta - \sum_b (\phi_{ab} \star L_b)(\theta)} \,, \\
    \\
    \displaystyle{L_a = \log \left(1+e^{-\epsilon_a} \right) \,,
    \quad \phi_{ab} = \frac{N_{ab}}{2\pi \cosh \theta} } \,, \\
    \\
    \displaystyle{E(\mu,R) = -\sum_a \frac{\mu_a}{2\pi} \int L_a(\theta) \cosh \theta \ d\theta} \,,
  \end{array}
\end{equation}
where $\mu_a$ is the mass of particles of type $a$, and $N_{ab}$ is the
adjacency matrix of a diagram describing the scattering between particles. In a diagonal
(non-reflecting) scattering
theory, the $\mu_a$ and $\phi_{ab}$ would be given directly from the dispersion relations
and the $S$-matrix for elementary excitations. However, the present model does allow reflection
of the particles. The main difficulty here is then to find the correct TBA
diagram and masses for the $S$-matrix we want to study.
Following the ideas of Refs.~\cite{TBA-RSOS,TBA-massless,FSZ} (see the Introduction), we conjecture that
the TBA diagram for the scattering theory~\eqref{eq:Smain} with mass
$\mu$~\eqref{eq:mass} is the diagram of Fig.~\ref{fig:TBA-RSOS}.

Now, assuming the above conjecture is correct, we check that the TBA
equations~\eqref{eq:TBA} for the diagram of Figure~\ref{fig:TBA-RSOS} lead to
the central charge~\eqref{eq:cc} in the UV limit $R \to 0$.
As in Ref.~\cite{TBA-RSOS}, the ground-state energy in the UV limit is obtained in terms of
the limiting values of $\epsilon_a$ in the UV and IR limit:
\begin{equation} \label{eq:E-UV}
  E(\mu,R) \simeq -\frac{1}{\pi R}  \left[
    \sum_{a=1}^{t-3} {\cal L} \left( \frac{x_a}{1+x_a} \right)
    - \sum_{a=1}^{t-3} {\cal L} \left( \frac{y_a}{1+y_a} \right)
  \right] \,,
  \qquad x_a= \lim_{R \to 0} \left[\rme^{-\epsilon_a(0)} \right] \,,
  \quad y_a= \lim_{R \to \infty} \left[\rme^{-\epsilon_a(0)} \right] \,,
\end{equation}
where $\cal L$ is the Rogers dilogarithm:
\begin{equation}
  {\cal L}(x) = -\half \int_0^x dt \left[ \frac{\log t}{1-t} + \frac{\log(1-t)}{t} \right] \,.
\end{equation}
The quantities $x_a,y_a$ are determined by the adjacency matrix $N_{ab}$ and
the masses $\mu_a$~\cite{TBA-RSOS}:
\begin{equation}
  x_a^2 = \prod_{b=1}^{t-3} (1+x_b)^{N_{ab}} \,,
  \qquad y_a^2 = \prod_{b | \mu_b=0} (1+y_b)^{N_{ab}} \,.
\end{equation}
To connect this with known results~\cite{TBA-RSOS} on the RSOS central charge, we introduce
the quantities $z_a$ which satisfy:
\begin{equation}
  z_a^2 = \prod_{b=2}^{t-3} (1+z_b)^{N_{ab}} \,, \qquad a=2,\dots,t-3 \,,
\end{equation}
and write~\eqref{eq:E-UV} as:
\begin{equation}
  E(\mu,R) \simeq -\frac{1}{\pi R} \left[
      \sum_{a=1}^{t-3} {\cal L} \left( \frac{x_a}{1+x_a} \right)
      - \sum_{a=2}^{t-3} {\cal L} \left( \frac{z_a}{1+z_a} \right)
    \right]
    -\frac{1}{\pi R} \left[
      \sum_{a=2}^{t-3} {\cal L} \left( \frac{z_a}{1+z_a} \right)
      - \sum_{a=2}^{t-4} {\cal L} \left( \frac{y_a}{1+y_a} \right)
    \right] \,.
\end{equation}
The above expression is exactly the sum of ground-state energies for the
$A_{t-2}$ and $A_{t-3}$ RSOS models, so the central charge is:
\begin{equation} \label{eq:cc-UV}
  c = \left[ 1-\frac{6}{t(t-1)} \right] 
  + \left[ 1-\frac{6}{(t-1)(t-2)} \right]
  = 2-\frac{12}{t(t-2)} \,,
\end{equation}
which is the central charge~\eqref{eq:cc} of the critical theory.

\begin{figure}
  \begin{center}
    \includegraphics{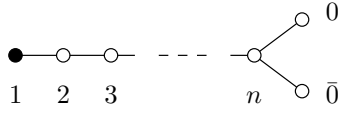}
  \end{center}
  \caption{The TBA for the Sine-Gordon model at coupling 
    $\til{\beta}_{\rm SG}^2/(8\pi)=(n+1)/(n+2)$.}
  \label{fig:tbaSG}
\end{figure}

Finally, we show that, throughout the scaling regime, the ground-state energy
$E(\mu,R)$ is twice that of the corresponding twisted Sine-Gordon
model. Different cases arise, and we will discuss only one: the case when
$t-3=2n+1, n \in \mathbb{N}$. We can then relabel the nodes on the diagram of
Figure~\ref{fig:TBA-RSOS}, so that the $n$ leftmost ones are called
$1,\ldots,n$, the $n$ rightmost ones $\bar{n},\ldots,\bar{1}$, and the middle
one $0$. The TBA equations~\eqref{eq:TBA} then read:
\begin{eqnarray}
  \epsilon_a &=& \delta_{a1} \mu R\cosh\theta 
  - \sum_b N_{ab} \ \phi \star L_b \,,
  \qquad a=1,\ldots,n-1 \nonumber\\
  \epsilon_{\bar{a}} &=& \delta_{\bar{a}\bar{1}} \mu R\cosh\theta
  -\sum_b N_{ab} \ \phi \star L_{\bar{b}} \,,
  \qquad a=1,\ldots,n-1 \nonumber\\
  \epsilon_{n} &=& -\phi \star \left(L_{n-1}+L_{0}\right) \nonumber\\
  \epsilon_{\bar{n}} &=& -\phi \star \left(L_{\overline{n-1}}+L_{0}\right) \nonumber\\
  \epsilon_{0} &=& -\phi \star \left(L_{n}+L_{\bar{n}}\right),
\end{eqnarray}
where we have set $\phi(\theta) = 1/(2\pi \cosh \theta)$.
We consider symmetric solutions under the exchange of $a$ and $\bar{a}$:
\begin{eqnarray}
  \epsilon_a &=& \delta_{a1} \mu R\cosh\theta - \sum_b N_{ab} \phi\star L_b \,,
  \qquad a=1,\ldots,n-1 \nonumber\\
  \epsilon_{n} &=& -\phi \star L_{n-1} 
  - \phi \star \log \left( 1 + \rmi \ \rme^{-\epsilon_0/2} \right)
  - \phi \star \log \left( 1 - \rmi \ \rme^{-\epsilon_0/2} \right) \nonumber\\
  \frac{\epsilon_{0}}{2} &=& -\phi \star L_{n} \nonumber\\
  \epsilon_{\bar{a}} &=& \epsilon_a \,.
\end{eqnarray}
The ground state energy is meanwhile:
\begin{equation}
  E(\mu,R) = -2\sum_{a=1}^n \frac{\mu_a}{2\pi} \int L_a(\theta) \cosh\theta \ d\theta \,.
\end{equation}
We thus see that our system has twice the ground-state energy of a TBA whose diagram is as in 
Figure~\ref{fig:tbaSG}, and which involves a fugacity for the two end nodes of the fork equal
to $\pm \rmi$. This is exactly the TBA for the twisted Sine-Gordon model, following the lines
of Ref.~\cite{FS}. For this value of the twist in particular, the  results in Ref.~\cite{FS} give the 
central charge (using eq.(21) of~\cite{FS}, with $t+1=n+2$ the total number of nodes;
$\lambda_{1^-}=-\lambda_{t-1}=\rmi$ in eq.(20) of~\cite{FS}):
\begin{equation} \label{eq:cc-SG}
  \til{c} = 1-\frac{3/2}{(n+1)(n+2)} \,.
\end{equation}
Comparing \eqref{eq:cc-UV} and \eqref{eq:cc-SG} for $t-3=2n+1$, we see that $c = 2 \til{c}$.
 
\subsection{The field theory}
 
We now try to identify the field theory described by our TBA. This of course involves a bit of guesswork. 
 
First, recall we have found that the physical mass of the theory scales as $\mu \propto e^{-t\Lambda/2}$ where
$\gamma = \pi/t$. On the other hand, we can in general expect that we are facing the  perturbation of a 
model of central charge $c=2$ by some operator $\Phi_h$ of conformal dimension $h$, with action:
\begin{equation}
  A=A_{\rm CFT} + \omega \int d^2x \ \Phi_h \,,
\end{equation}
where $A_{\rm CFT}$ is the action for the critical UV limit.
The dimension of the coupling constant $\omega$ is $[\omega]=R^{2h-2}$, and thus the mass in the TBA scales as 
$\mu \propto \omega^{1/(2-2h)}$. A detailed look at the microscopic Hamiltonian shows that the bare coupling
is proportional to $\rme^{-\Lambda}$. If follows that $2-2h=2/t$, and thus: 
\begin{equation}
  h = \frac{t-1}{t} \,.
\end{equation}

In general, the TBA approach will give for the ground-state energy $E(\mu,R)$ a series in $\mu^\alpha$,
where the exponent $\alpha$ is given by $\alpha=2(1-h)$ if correlations of $\Phi_h$ are non zero 
({\it e.g.} perturbation by $\Phi_{13}$ in minimal models), $\alpha=4(1-h)$ if only even correlators
are non zero ({\it e.g.} the Sine-Gordon model). Other possibilities exist, {\it e.g.} 
$\alpha=8(1-h)$ if only correlators involving a number of operators 
{\it multiple of four} are non zero. 

Since the ground-state energy of our model is twice that of the SG model at 
$\til{\beta}_{\rm SG}^2/(8\pi)=(t-2)/t$, it follows that
$\alpha=4(1-h_{\rm SG})$, where $h_{\rm SG}=(t-2)/t$. Meanwhile, we have identified earlier the 
dimension of the perturbation as $h=(t-1)/t$, and thus $\alpha=8(1-h)$. We are forced to conclude therefore that in our problem indeed, {\it only correlators involving a number of operators multiple of four are non zero}. 

Meanwhile, the structure of the scattering matrix suggests the same quantum group symmetry as the one in the $\beta_{\rm SG}$ theory, with, for generic values of $t$, only one conserved charge, since the $S^{01}$ elements allow reflection of charges between $0$ and $1$ sectors (this does not occur at the special points where $t/(t-2)$ is an integer). Finally, the structure of finite-size effects showed that the CFT was made of a Dirac fermion and a boson of $t$-dependent radius. This all leads us to propose that the action of the theory is:
\begin{equation} \label{eq:big-guess}
  A[\varphi,\psi_1,\psi_2] = \int \left[ \half \partial_\nu \varphi \partial_\nu \varphi
    + \rmi(\bar{\psi}_1 \dbar \psi_1 + \bar{\psi}_2 \dbar \psi_2) \right] d^2x
  + \mu_0 \int \left[ \psi_1\bar{\psi}_1 \rme^{\rmi\beta\varphi} + \psi_2\bar{\psi}_2 \rme^{-\rmi\beta\varphi}
  \right] d^2x \,,
\end{equation}
where $\varphi$ is the boson, $\psi_1,\psi_2$ are the two Majorana components of the Dirac theory.
One can check that this theory is indeed integrable using the non-local conserved charges
$\psi_1 \rme^{-(4\rmi\pi/\beta)\varphi}$ and $\psi_2 \rme^{(4\rmi\pi/\beta)\varphi}$. The algebra satisfied by these
charges leads to $\SU(2)_q$ with quantum-group deformation parameter \cite{SSim}:
\begin{equation}
  q = -\exp(-\rmi\pi/\delta) \,, 
  \qquad \delta = \frac{2\beta^2}{4\pi-\beta^2} \,.
\end{equation}
Meanwhile, the basic SG $S$-matrix with the foregoing value of $\beta_{\rm SG}$ has also quantum group 
symmetry~\cite{SG-q-group}, with deformation parameter that corresponds to:
\begin{equation}
  \delta = \frac{\beta_{SG}^2}{8\pi-\beta_{SG}^2}=t-2 \,.
\end{equation}
By requiring that the symmetry of the $S$-matrix is the symmetry of the action, we identify
the two above expressions for $\delta$, and we get:
\begin{equation}
  \frac{\beta^2}{8\pi} = \frac{t-2}{2t} \,.
\end{equation}
The dimension of the perturbation is indeed:
\begin{equation}
  h= \half + \frac{t-2}{2t}= \frac{t-1}{t} \,,
\end{equation}
and clearly only correlators involving a number of operators multiple of four are non zero. 
We also obtain a non-unitary theory, which is expected from the presence of complex terms in the
Hamiltonian. 

An important check of our proposal would be to see if the ground-state  energy of theory \eqref{eq:big-guess}
is  twice the ground-state energy of the related Sine-Gordon model. One might first tackle this
question in perturbation theory. We will leave this for future work, and content ourselves by examining 
the question in the limit $\beta \to 0$. Then $\til{\beta}_{\rm SG}\to 0$ and we expect, on
the one hand, the ground-state energy to be twice the one of a free boson. On the other hand, our action
reduces naively to two identical massive Majorana fermions. In the limit $\beta \to 0$  however,
counter-terms are needed and a $\varphi^2$ term also appears (exactly as in the case of $N=1$ theories
\cite{Takacs}), leading to an additional free massive boson. Denote $E_b$ the ground-state energy
 of such a boson, and $E_f$ the ground-state energy of a free Majorana fermion:
\begin{eqnarray}
  E_b(\mu,R) &=& -\frac{\mu}{2\pi} \int_{-\infty}^\infty \log \left( 1- \rme^{-\mu R \cosh\theta} \right)
  \cosh \theta \ d\theta \,, \nonumber \\
  E_f(\mu,R) &=&  \frac{\mu}{2\pi} \int_{-\infty}^\infty \log \left( 1+ \rme^{-\mu R \cosh\theta} \right)
  \cosh \theta \ d\theta \,.
\end{eqnarray}
We have the identity
\begin{equation} \label{result}
  2E_b(\mu,R) = E_b(2\mu, R) + 2E_f(\mu,R) \,,
\end{equation}
so we see indeed that the ground-state energy of our field theory will be twice the ground-state
energy of the SG model in the limit of vanishing coupling provided the mass terms are in the proper ratios. 
More precisely, the left-hand side of~\eqref{result} corresponds to twice the ground-state energy for
the theory:
\begin{equation}
  A[\varphi] = \int \left( \half \partial_\nu \varphi \partial_\nu \varphi+ \mu^2 \varphi^2 \right) \,,
\end{equation}
so near $\beta_{\rm SG}=0$ we will need an action of  the form:
\begin{equation} \label{eq:action}
  A[\varphi,\psi_1,\psi_2] = \int \left[ \half \partial_\nu \varphi \partial_\nu \varphi
    + \rmi(\bar{\psi}_1 \dbar \psi_1 + \bar{\psi}_2 \dbar \psi_2) \right] d^2x
  + \mu_0 \int \left[ \psi_1\bar{\psi}_1 \rme^{\rmi\beta\varphi} + \psi_2\bar{\psi}_2 \rme^{-\rmi\beta\varphi}
  \right] d^2x
  + \frac{\mu_0^2}{\beta^2} \int \cos(2\beta\varphi) d^2x \,.
\end{equation}
We now observe that our theory is identical to the $C_2^{(2)}$ Toda theory (more precisely, we need in
fact to set $n=1$ in the more general $C_{n+1}^{(2)}$ theory, whose form is valid for $n>1$ only),
whose Lagrangian would read \cite{DFdV,DGPZ}:
\begin{equation}
  {\cal L}[\varphi,\psi_1,\psi_2] = \half \partial_\nu \varphi \partial_\nu \varphi
  +\rmi(\bar{\psi}_1 \dbar \psi_1 + \bar{\psi}_2 \dbar \psi_2)
  -\mu_0 (\bar{\psi}_1 \psi_1 \rme^{g\varphi/\sqrt{2}} + \bar{\psi}_2\psi_2 \rme^{-g\varphi/\sqrt{2}})
  - \frac{2\mu_0^2}{g^2} \cosh(\sqrt{2}g\varphi) \,.
\end{equation}
Clearly we have to set $g=\rmi\sqrt{2}\beta$. We then see that in the limit $\beta\rightarrow 0$ 
the boson has mass parameter twice the one of the Majorana fermions, in agreement with
\eqref{result}. To summarise our results:
\smallskip
\begin{itemize}
\item The continuum limit of our lattice model is the 
  complex $C_2^{(2)}$ theory~\eqref{eq:action}. 
\item The $S$-matrix of this theory is given by \eqref{eq:Smain},
  with $\beta_{\rm SG}^2/(8\pi)=2\beta^2/(\beta^2+4\pi)$.
\item The ground-state energy is twice the ground-state energy of the Sine-Gordon theory
  at coupling $\til{\beta}_{\rm SG}^2=2\beta^2$.
\end{itemize}

\section{Conclusion}

From the point of view of integrable statistical models, one can think of
several ways to generalise the construction of the $\Zbb_2$ model.  First, one
can build a model with a staggering of period $n>2$, which has a $\Zbb/n\Zbb$
symmetry~\cite{these}.  We guess that the corresponding continuum limit will
be related to a product of $n$ copies of $\SU(2)_1$, made anisotropic by a
$J^3J^3$ term as in the case $n=2$. What the integrable massive deformation
might be is however more mysterious.  Also, the effective field theory for the
analog of the non-compact regime~\cite{IJS-PAF} is less clear.  Another
interesting direction is to apply the same kind of construction to other
models than the six-vertex model. Of particular interest here would be the
`dilute version', obtained by staggering the Izergin-Korepin 19-vertex model.

In the CFT perspective, the expression for the toroidal partition function in
terms of Coulombic partition functions generally leads to a classification of
new minimal series of CFTs. It is possible, in principle,
to follow this program in the case of the $\Zbb_2$ model partition functions.

There are also some important questions about the physical interpretation of
the $\Zbb_2$ Hamiltonian as a zig-zag spin chain. We have seen that
non-Hermitian terms in the Hamiltonian play a role, but it could be that the
model is in the same universality class as a well-defined, Hermitian
spin-chain model. Additionnally, at the Majumdar-Ghosh point, the gapped
excitations above the ground state (spinons) could be studied more
systematically, through a variational approach similar to~\cite{MG-spinons}.

In the context of anyonic chains, the RSOS version of the $\Zbb_2$ model
is an integrable point in the phase diagram of the three-anyon interaction
Hamiltonian~\eqref{eq:H-anyons2}. It actually governs the behaviour of this system throughout a
whole critical phase, as was shown numerically in Ref.~\cite{IJS-TL}. Various
features of this phase diagram still lead to open questions, such as the complete RG flow
of~\eqref{eq:H-anyons2} and the associated operators at the fixed points, but also
the differences between the RSOS and loop formulations.

\subsection*{Acknowledgments}
The authors thank Paul Fendley for clarifications on the link between RSOS
models and anyonic fusion rules.
YI thanks Steve Simon and Eddy Ardonne for useful
discussions on anyons, and Fabian Essler for comments on spin chains.
The work of JLJ was supported by the European Community Network ENRAGE (grant
MRTN-CT-2004-005616); that of HS by the ESF Network INSTANS; and that of
JLJ and HS by the Agence Nationale de la Recherche (grant
ANR-06-BLAN-0124-03).

\section*{Appendix A: Physical quantities for holes}
\renewcommand\thesection{A}
\setcounter{equation}{0}

This Appendix is about the analysis of the Bethe equations in the continuum limit.
Here we prove Eq.~\eqref{eq:dA}, which gives the variation of a physical quantity $A$
in the presence of a hole $\lambda_h$ in the distribution $\{\lambda_j^{(0)}\}$.
It is useful first to give the Fourier transform of the momentum and the
kernels:
\begin{equation}
  2\widehat{k'}(\omega) = \frac{2\pi \sinh (\pi/2-\gamma)\omega}{\sinh (\pi \omega/2)} \,,
\end{equation}
\begin{equation}
  \Khat^{(0)}(\omega) = 
  -\frac{2 \pi \sinh (\pi-2\gamma) \omega}{\sinh \pi \omega} \,,
  \qquad \Khat^{(\pm 1)}(\omega) = 
  \frac{2 \pi \sinh 2 \gamma \omega}{\sinh \pi \omega} \,,
\end{equation}
\begin{equation} \label{eq:J-pm}
  1 + \Jhat^{(+)}(\omega) = \frac{\sinh(\pi \omega/2)}
  {2 \sinh(\pi/2-\gamma)\omega \ \cosh \gamma \omega} \,,
  \qquad 1 + \Jhat^{(-)}(\omega) = \frac{\cosh(\pi \omega/2)}
  {2 \cosh(\pi/2-\gamma)\omega \ \cosh \gamma \omega} \,.
\end{equation}
The hole $\lambda_h$ affects the Lieb equations~\eqref{eq:lieb}:
\begin{equation}
  \left\{ \begin{array}{rcl}
    2 k'(\lambda) &=& [(2\pi-K^{(0)}) \star \rho^{(0)}](\lambda) - (K^{(1)} \star \rho^{(1)})(\lambda)
    + \frac{1}{N} K^{(0)}(\lambda-\lambda_h) \,, \\
    \\
    2 k'(\lambda) &=& - (K^{(1)} \star \rho^{(0)})(\lambda) + [(2\pi-K^{(0)}) \star \rho^{(1)}](\lambda) 
    + \frac{1}{N} K^{(1)}(\lambda-\lambda_h) \,,
  \end{array} \right.
\end{equation}
Combining with the ground-state equation, we get:
\begin{equation}
  \rho^{(0)}(\lambda)+\rho^{(1)}(\lambda)-2\rho_\infty(\lambda)
  = -\frac{1}{N} J^{(+)}(\lambda-\lambda_h) \,.
\end{equation}
The variation of $A$ is given by:
\begin{eqnarray}
  A-A_0 &=& \int_{-\infty}^\infty d\lambda \ [\rho^{(0)}(\lambda)-\rho_\infty(\lambda)] \ \alpha(\lambda)
  + \int_{-\infty}^\infty d\lambda \ [\rho^{(1)}(\lambda)-\rho_\infty(\lambda)] \ \alpha(\lambda)
  - \frac{1}{N} \alpha(\lambda_h) \nonumber \\
  &=& \int_{-\infty}^\infty d\lambda (\rho^{(0)}+\rho^{(1)}-2\rho_\infty)(\lambda) \ \alpha(\lambda)
  - \frac{1}{N} \alpha(\lambda_h) \nonumber \\
  &=& - \frac{1}{N} \int_{-\infty}^\infty d\lambda \ J^{(+)}(\lambda-\lambda_h) \ \alpha(\lambda) 
  - \frac{1}{N} \alpha(\lambda_h)\,.
\end{eqnarray}
Since $J^{(+)}$ is even, we get the result~\eqref{eq:dA}.

\section*{Appendix B: Finite-size corrections}
\renewcommand\thesection{B}
\setcounter{equation}{0}

In this Appendix, we introduce a variant of the Wiener-Hopf method~\cite{YY},
to calculate finite-size corrections to the energies from the analysis of the Bethe equations.

We consider combined magnetic excitations $(m^{(0)},m^{(1)})$.
Since the Bethe integer distributions $\{ I_j^{(0,1)}\}$ are symmetric around zero, the bounds of the integrals
in Eq.~\eqref{eq:lieb} are such that $C^{(0)}=B^{(0)}, C^{(1)}=B^{(1)}$. We can write:
\begin{equation}
  \left\{ \begin{array}{rcl}
    2k'(\lambda) &=& \displaystyle{
      2\pi \rho^{(0)}(\lambda)
      - \int_{-B^{(0)}}^{+B^{(0)}} d\mu \ \rho^{(0)}(\mu) K^{(0)}(\lambda-\mu)
      - \int_{-B^{(1)}}^{+B^{(1)}} d\mu \ \rho^{(1)}(\mu) K^{(-1)}(\lambda-\mu)
    } \\
    2k'(\lambda) &=& \displaystyle{
      2\pi \rho^{(1)}(\lambda)
      - \int_{-B^{(0)}}^{+B^{(0)}} d\mu \ \rho^{(0)}(\mu) K^{(1)}(\lambda-\mu)
      - \int_{-B^{(1)}}^{+B^{(1)}} d\mu \ \rho^{(1)}(\mu) K^{(0)}(\lambda-\mu)
    } \\
    2k'(\lambda) &=& \displaystyle{
      2\pi \rho_\infty(\lambda)
      - \int_{-\infty}^{\infty} d\mu \ \rho_\infty(\mu) K^{(0)}(\lambda-\mu)
      - \int_{-\infty}^{\infty} d\mu \ \rho_\infty(\mu) K^{(-1)}(\lambda-\mu)
    } \,.
  \end{array} \right.
\end{equation}
Applying the convolution by $(2\pi)^{-1} (\delta+J^{(\pm)})$ to symmetric and antisymmetric combinations,
we get:
\begin{equation} 
  \left\{ \begin{array}{rcl}
    (\rho^{(0)}+\rho^{(1)}-2\rho_\infty)(\lambda)
    &=& \displaystyle{
      - \int_{|\mu|>B^{(0)}} d\mu \ \rho^{(0)}(\mu) J^{(+)}(\lambda-\mu)
      - \int_{|\mu|>B^{(1)}} d\mu \ \rho^{(1)}(\mu) J^{(+)}(\lambda-\mu)
    } \\
    (\rho^{(0)}-\rho^{(1)})(\lambda)
    &=& \displaystyle{
      - \int_{|\mu|>B^{(0)}} d\mu \ \rho^{(0)}(\mu) J^{(-)}(\lambda-\mu)
      + \int_{|\mu|>B^{(1)}} d\mu \ \rho^{(1)}(\mu) J^{(-)}(\lambda-\mu) 
    } \,.
  \end{array} \right.
\end{equation}
Combining again the equations, we get:
\begin{equation} \label{eq:lieb-rho}
  \left\{ \begin{array}{rcl}
    (\rho^{(0)}-\rho_\infty)(\lambda)
    &=& \displaystyle{
      - \int_{|\mu|>B^{(0)}} d\mu \ \rho^{(0)}(\mu) J^{(0)}(\lambda-\mu)
      - \int_{|\mu|>B^{(1)}} d\mu \ \rho^{(1)}(\mu) J^{(-1)}(\lambda-\mu) 
    } \\
    (\rho^{(1)}-\rho_\infty)(\lambda)
    &=& \displaystyle{
      - \int_{|\mu|>B^{(0)}} d\mu \ \rho^{(0)}(\mu) J^{(1)}(\lambda-\mu)
      - \int_{|\mu|>B^{(1)}} d\mu \ \rho^{(1)}(\mu) J^{(0)}(\lambda-\mu) \,.
    }
  \end{array} \right.
\end{equation}
where $J^{(0)},J^{(\pm 1)}$ are defined in~\eqref{eq:scatt-amp}.
Let us define the symmetric/antisymmetric physical quantities:
\begin{equation}
  A^{(\pm)} \equiv \int_{-B^{(0)}}^{+B^{(0)}} d\lambda \ \rho^{(0)}(\lambda) \alpha(\lambda)
  \pm \int_{-B^{(1)}}^{+B^{(1)}} d\lambda \ \rho^{(1)}(\lambda) \alpha(\lambda) \,.
\end{equation}
The variation of $A^{(\pm)}$ with respect to the ground-state value $(A^{(\pm)})_\infty$
can be expressed as:
\begin{equation}
  A^{(\pm)} - A^{(\pm)}_\infty
  = -\left[ \int_{|\mu|>B^{(0)}} d\mu \ \rho^{(0)}(\mu) 
    [(\delta + J^{(\pm)}) \star \alpha](\mu)
    \pm \int_{|\mu|>B^{(1)}} d\mu \ \rho^{(1)}(\mu) 
    [(\delta + J^{(\pm)}) \star \alpha](\mu)
  \right]\,,
\end{equation}
where we use the fact that $J^{(+)}$ and $J^{(-)}$ are even.
Setting $\alpha(\lambda)= -1$ or $\alpha(\lambda) = \epsilon(\lambda)$, we get the charges 
and the energy:
\begin{eqnarray}
  \frac{m}{N} &=& [1+\Jhat^{(+)}(0)] \left( 
    \int_{|\mu|>B^{(0)}} d\mu \ \rho^{(0)}(\mu)
    + \int_{|\mu|>B^{(1)}} d\mu \ \rho^{(1)}(\mu) 
  \right)\,, \\
  \frac{\mt}{N} &=& [1+\Jhat^{(-)}(0)] \left( 
    \int_{|\mu|>B^{(0)}} d\mu \ \rho^{(0)}(\mu)
    - \int_{|\mu|>B^{(1)}} d\mu \ \rho^{(1)}(\mu) 
  \right)\,, \\
  \frac{E - E_{\rm gs}}{N} &=& \int_{|\mu|>B^{(0)}} d\mu \ \rho^{(0)}(\mu) \epsilon_d(\mu)
  +\int_{|\mu|>B^{(1)}} d\mu \ \rho^{(1)}(\mu) \epsilon_d(\mu)\,.
\end{eqnarray}

To solve the Lieb equations~\eqref{eq:lieb-rho}, we define 
the shifted densities: $g^{(a)}(\lambda) = \rho^{(a)}(B^{(a)}+\lambda)$ for $a=0,1$. 
Neglecting the terms from $\mu<0$ (see~\cite{YY}), we get the
coupled Wiener-Hopf equations:
\begin{equation}
  \left\{ \begin{array}{rcl}
    \displaystyle{
      g^{(0)}(\lambda) 
      + \int_0^\infty d\mu \ g^{(0)}(\mu) J^{(0)}(\lambda-\mu)
      + \int_0^\infty d\mu \ g^{(1)}(\mu) J^{(-1)}(\lambda-\mu+b) }
    &=& \rho_\infty(B^{(0)}+\lambda) \\
    \displaystyle{
      g^{(1)}(\lambda) 
      + \int_0^\infty d\mu \ g^{(0)}(\mu) J^{(1)}(\lambda-\mu-b)
      + \int_0^\infty d\mu \ g^{(1)}(\mu) J^{(0)}(\lambda-\mu) }
    &=& \rho_\infty(B^{(1)}+\lambda) \,,
  \end{array} \right.
\end{equation}
where $b=B^{(0)}-B^{(1)}$. After Fourier transform:
\begin{equation} \label{eq:wh}
  \left\{ \begin{array}{rcl}
    \big[ 1+\Jhat^{(0)}(\omega) \big] \ g^{(0)}_+(\omega) 
    + \rme^{-\rmi \omega b} \Jhat^{(-1)}(\omega) \ g^{(1)}_+(\omega) + g^{(0)}_-(\omega)
    &=& \rme^{-\rmi\omega B^{(0)}} \rhohat_\infty(\omega)  \\
    \rme^{\rmi\omega b} \Jhat^{(1)}(\omega) \ g^{(0)}_+(\omega)
    + \big[ 1+\Jhat^{(0)}(\omega) \big] \ g^{(1)}_+(\omega) 
    + g^{(1)}_-(\omega)
    &=& \rme^{-\rmi\omega B^{(1)}} \rhohat_\infty(\omega) \,.
  \end{array} \right.
\end{equation}
We use the factorisations:
\begin{equation}
  1+ \Jhat^{(+)}(\omega) = \frac{1}{G_+(\omega) G_-(\omega)} \,, \quad
  1+ \Jhat^{(-)}(\omega) = \frac{1}{H_+(\omega) H_-(\omega)}\,,
\end{equation}
where:
\begin{eqnarray}
  G_+(\omega) = \sqrt{\frac{2\pi^2}{\pi-2\gamma}} \ \frac{\Gamma(\rmi \omega / 2)}
  {\Gamma[ (1/2-\gamma/\pi) \rmi \omega ] \ \Gamma[1/2+ (\gamma/\pi) \rmi \omega]} \,,
  &\qquad& G_-(\omega) = G_+(-\omega) \,, \\
  H_+(\omega) = \frac{\sqrt{2\pi} \ \Gamma(1/2+ \rmi \omega/2)}
  {\Gamma[1/2 + (1/2-\gamma/\pi)\rmi \omega] \ \Gamma[1/2+ (\gamma/\pi) \rmi \omega]} \,,
  &\qquad& H_-(\omega) = H_+(-\omega) \,,
\end{eqnarray}
and we factorize the $2 \times 2$ matrix:
\begin{equation}
  {\bf 1} + {\bf \Jhat} \equiv \left( \begin{array}{cc}
      1+\Jhat^{(0)} & \rme^{-\rmi\omega b} \Jhat^{(-1)} \\
      \rme^{\rmi\omega b} \Jhat^{(1)} & 1+\Jhat^{(0)}
    \end{array} \right)
  = {\bf G}_-^{-1} \ {\bf G}_+^{-1}\,.
\end{equation}
The matrices ${\bf G}_\pm$ read:
\begin{eqnarray}
  {\bf G}_\pm &=& \half \left( \begin{array}{cc}
      G_\pm+H_\pm & \rme^{-\rmi \omega b}(G_\pm-H_\pm) \\
      \rme^{\rmi \omega b}(G_\pm-H_\pm) & G_\pm+H_\pm
    \end{array} \right) \,, \\
  {\bf G}_\pm^{-1} &=& \half \left( \begin{array}{cc}
      G_\pm^{-1}+H_\pm^{-1} & \rme^{-\rmi \omega b}(G_\pm^{-1}-H_\pm^{-1}) \\
      \rme^{\rmi \omega b}(G_\pm^{-1}-H_\pm^{-1}) & G_\pm^{-1}+H_\pm^{-1}
    \end{array} \right) \,.
\end{eqnarray}
We can write the system~\eqref{eq:wh} as:
\begin{equation}
  ({\bf 1} + {\bf \Jhat}) \left( \begin{array}{c}
      g^{(0)}_+ \\
      g^{(1)}_+ \end{array} \right)
  + \left( \begin{array}{c}
      g^{(0)}_- \\
      g^{(1)}_- \end{array} \right)
  = \rme^{-\rmi\omega B^{(0)}} \rhohat_\infty \left( \begin{array}{c}
      1 \\
      \rme^{\rmi\omega b} \end{array} \right)\,.
\end{equation}
We multiply by ${\bf G}_-$:
\begin{equation}
  {\bf G}_+^{-1} \left( \begin{array}{c}
      g^{(0)}_+ \\
      g^{(1)}_+ \end{array} \right)
  + {\bf G}_- \left( \begin{array}{c}
      g^{(0)}_- \\
      g^{(1)}_- \end{array} \right)
  = \rme^{-\rmi \omega B^{(0)}} \rhohat_\infty \ {\bf G}_- \left( \begin{array}{c}
      1 \\
      \rme^{\rmi \omega b} \end{array} \right)\,.
\end{equation}
The solution is given in terms of the pole $\omega_0= -\rmi\pi/(2\gamma)$ for $\rhohat_\infty$ 
and the residue $r_0={\rm Res}(\rhohat_\infty, \omega_0)$:
\begin{eqnarray}
  \left( \begin{array}{c}
      g^{(0)}_+ \\
      g^{(1)}_+ \end{array} \right)
  &=& {\bf G}_+ \left[ \rme^{-\rmi \omega B^{(0)}} \rhohat_\infty
    \ {\bf G}_- \left( \begin{array}{c}
        1 \\
        \rme^{\rmi \omega b} \end{array} \right) \right]_+ \nonumber \\
  &=& -\frac{r_0 \zeta^{(0)}}{\omega_0-\omega}
  {\bf G}_+(\omega) {\bf G}_-(\omega_0) \left( \begin{array}{c}
      1 \\
      \rme^{\rmi \omega_0 b} \end{array} \right) \nonumber \\
  &=& -\frac{r_0 \zeta^{(0)} G_-(\omega_0)}{\omega_0-\omega}
  {\bf G}_+(\omega) \left( \begin{array}{c}
      1 \\
      \rme^{\rmi \omega_0 b} \end{array} \right)\,,
\end{eqnarray}
where $\zeta^{(a)} = \rme^{-\rmi \omega_0 B^{(a)}}$. So the magnetic charges are given by:
\begin{eqnarray}
  \frac{m}{N} &=& -\frac{2 r_0 G_-(\omega_0)}{\omega_0 G_-(0)} (\zeta^{(0)}+\zeta^{(1)})\,, \\
  \frac{\mt}{N} &=& -\frac{2 r_0 G_-(\omega_0)}{\omega_0 H_-(0)} (\zeta^{(0)}-\zeta^{(1)})\,.
\end{eqnarray}
The total energy is:
\begin{eqnarray}
  \frac{E-E_{\rm gs}}{N} &=&
  2\rmi \ {\rm Res} (\widehat{\epsilon}_d, \omega_0)
  \ [ \zeta^{(0)} g_+^{(0)}(-\omega_0) + \zeta^{(1)} g_+^{(1)}(-\omega_0) ]
  \nonumber \\
  &=& 2\rmi \pi \sin 2\gamma
  \ \frac{r_0^2 G_-(\omega_0) G_+(-\omega_0)}{\omega_0} [ (\zeta^{(0)})^2+(\zeta^{(1)})^2 ] \nonumber \\
  &=& \frac{2 \pi v}{8} \left[ 
    \left(\frac{G_-(0) m}{N}\right)^2 
    + \left(\frac{H_-(0) \mt}{N}\right)^2 \right] \nonumber \\
  &=& \frac{2 \pi v}{8} \left\{
    [1+\Jhat^{(+)}(0)]^{-1} \left(\frac{m}{N}\right)^2
    + [1+\Jhat^{(-)}(0)]^{-1} \left(\frac{\mt}{N}\right)^2
  \right\} \,.
\end{eqnarray}
Using expressions~\eqref{eq:J-pm} for $\Jhat^{(\pm)}$, we get the critical exponents given
in \eqref{eq:Delta}. A similar calculation with $C^{(a)} \neq B^{(a)}$ would give the
electric critical exponents.

\section*{Appendix C: Bosonic partition functions and Jacobi's theta functions}
\renewcommand\thesection{C}
\setcounter{equation}{0}

\subsection*{Free boson on a torus}

Let us recall some known results on the free boson theory on a torus~\cite{CG}.
We denote by $\tau$ the modular ratio of the torus, and we write $q=\rme^{2\rmi \pi \tau}$.
The free boson is defined by the action $\cal A$ and the partition function~$Z_0$:
\begin{eqnarray}
  {\cal A}[\varphi] &=& \frac{g}{4\pi} \int d^2x \ |\nabla \varphi |^2  \,, \label{eq:A-boson} \\
  Z_0(g) &=& \int [D\varphi] \ \exp(-{\cal A}[\varphi]) 
  = \sqrt{\frac{g}{\im \tau}} \frac{1}{|\eta(\tau)|^2}  \,,
\end{eqnarray}
where $g$ is the coupling constant, and $\eta(\tau)$ is the Dedekind function:
\begin{equation} \label{eq:eta}
  \eta(\tau) = q^{1/24} \prod_{n=1}^\infty (1-q^n) \,.
\end{equation}
When defects $\delta \varphi, \delta' \varphi$ are introduced on the boundaries, this defines
the partition function~$Z_{m,m'}$, with $m,m'$ integers:
\begin{equation}
  Z_{m,m'}(g) = \int_{\myscript{\delta\varphi = 2\pi m}{\delta'\varphi = 2\pi m'}}
  [D\varphi] \ \exp(-{\cal A}[\varphi]) = Z_0(g) \ \exp \left( -\frac{\pi g |m'-m\tau|^2}{\im \tau} \right) \,. \label{eq:Zmm}
\end{equation}
A Poisson summation of~\eqref{eq:Zmm} yields:
\begin{equation} \label{eq:Poisson}
  \sum_{m' \in \Zbb} \rme^{\rmi \alpha m'} Z_{m,m'}(g) = \frac{1}{|\eta(\tau)|^2}
  \sum_{k \in \Zbb + \alpha/(2\pi)} q^{(k/\sqrt{g} + m\sqrt{g})^2/4} \ \qb^{(k/\sqrt{g} - m\sqrt{g})^2/4} \,.
\end{equation}

\subsection*{Jacobi theta functions}

The Jacobi theta functions are defined as:
\begin{equation} \label{eq:theta}
  \begin{array}{rclrcl}
    \theta_1(\tau) &=& {\displaystyle -\rmi \sum_{n \in \Zbb} (-1)^n q^{(n+1/2)^2/2}}=0 \,, & \qquad
    \theta_2(\tau) &=& {\displaystyle \sum_{n \in \Zbb} q^{(n+1/2)^2/2}} \,, \\
    \\
    \theta_3(\tau) &=& {\displaystyle \sum_{n \in \Zbb} q^{n^2/2}} \,, & \qquad
    \theta_4(\tau) &=& {\displaystyle \sum_{n \in \Zbb} (-1)^n q^{n^2/2}} \,.
  \end{array}
\end{equation}
They obey the algebraic relations:
\begin{eqnarray}
  \theta_2(\tau) \theta_3(\tau) \theta_4(\tau) &=& 2\eta(\tau)^3 \,, 
  \label{eq:theta-2tau} \\
  \sqrt{\theta_3(\tau) \theta_4(\tau)} &=& \theta_4(2\tau) \label{eq:Jacobi} \,.
\end{eqnarray}
We denote the Jacobi partition functions by:
\begin{equation} \label{eq:Znu}
  Z_\nu = \left| \frac{\theta_\nu(\tau)}{\eta(\tau)} \right| \,,
  \qquad \nu=2,3,4\,.
\end{equation}
The Ising partition functions $\Zc(r,r')$ are related to the Jacobi ones by:
\begin{equation} \label{eq:Z-Zc}
  \begin{array}{rcl}
    Z_2 &=& \Zc(1,0)+\Zc(1,1) \,, \\
    Z_3 &=& \Zc(0,1)+\Zc(1,0) \,, \\
    Z_4 &=& \Zc(0,1)+\Zc(1,1) \,, \\
    \Zc(0,0) &=& \Zc(0,1)+\Zc(1,0)+\Zc(1,1) \,.
  \end{array}
\end{equation}

\section*{Appendix D: Partition functions for the staggered models}
\renewcommand\thesection{D}
\setcounter{equation}{0}

\subsection*{Twisted vertex model and Potts model}

Starting from the untwisted partition function $Z(g)$, we can proceed like
in~\cite{CG}, to construct the twisted partition function and the Potts
partition function. The partition function where non-contractible loops
have a weight $\widehat{n} = 2\cos \phi$ is given by:
\begin{equation}
  \Zhat(g,\phi) = 2 \left(
    A \sum_{m,m' \ {\rm even}}
    + B \sum_{m \ {\rm even}, m' \ {\rm odd}}
    + C \sum_{m \ {\rm odd}, m' \ {\rm even}}
    + D \sum_{m,m' \ {\rm odd}}
  \right) Z_{m,m'}(g) \ \cos(2\phi \ m \wedge m') \label{pgcd} \,,
\end{equation}
where $m \wedge m'$ denotes the greatest common factor between $m$ and $m'$.
In particular, for $\phi= \pi/2, \pi/4$, we have:
\begin{eqnarray}
  \Zhat(g,\pi/2) &=& 2 \left(
    A \sum_{m,m' \ {\rm even}}
    - B \sum_{m \ {\rm even}, m' \ {\rm odd}}
    - C \sum_{m \ {\rm odd}, m' \ {\rm even}}
    - D \sum_{m,m' \ {\rm odd}}
  \right) Z_{m,m'}(g) \,, \\
  \Zhat(g,\pi/4) &=& A \left[
    \sum_{m,m' \in \Zbb} Z_{m,m'}\left(\frac{1}{16g}\right) - 2 \sum_{m,m' \in 2\Zbb} Z_{m,m'}(g)
  \right] \,.
\end{eqnarray}
The $Q$-state Potts partition function 
has an extra term due to clusters with cross geometry~\cite{CG}:
\begin{equation} \label{eq:ZPotts}
  Z_{\rm Potts}(Q) = \Zhat(g, \pi e_0) + \half (Q-1) \Zhat(g,\pi/2) \,,
\end{equation}
where
\begin{equation}
  \sqrt{Q}=2 \cos \gamma \,, 
  \qquad 0 < \gamma < \frac{\pi}{2} \,,
  \qquad g= \frac{\pi-2\gamma}{2\pi} \,, 
  \qquad e_0=\frac{\gamma}{\pi} \,.
\end{equation}

\subsection*{Particular values of $Q$}

\begin{itemize}
\item The case $Q=2$. \\
This provides a good check of the result~\eqref{eq:Zg-Ising},
since the Potts model arising from the staggered vertex model is equivalent,
on the lattice, to the usual critical Ising model. Using~\eqref{eq:ZPotts}:
\begin{equation}
  Z_{\rm Potts}(Q=2) = \left[
    (A-B) \sum_{m \ {\rm even}, m' \ {\rm odd}}
    +(A-C) \sum_{m \ {\rm odd}, m' \ {\rm even}}
    +(A-D) \sum_{m,m' \ {\rm odd}}
  \right] Z_{m,m'}(1/4) \,.
\end{equation}
Now the sums on $m,m'$ can be expressed in terms of the $Z_\nu$:
\begin{equation} \label{eq:sumZ4}
  \sum_{m \ {\rm even}, m' \ {\rm odd}} Z_{m,m'}(1/4) = \half Z_3 Z_4 \,, \quad
  \sum_{m \ {\rm odd}, m' \ {\rm even}} Z_{m,m'}(1/4)
  = \half Z_2 Z_3 \,, \quad
  \sum_{m,m' \ {\rm odd}} Z_{m,m'}(1/4)
  = \half Z_2 Z_4 \,.
\end{equation}
We obtained the first identity by using~\eqref{eq:theta-2tau}, and the two others by expanding the square of the left-hand sides. Combining~\eqref{eq:sumZ4} with \eqref{eq:ABCD} and \eqref{eq:Jacobi}, we get:
\begin{equation}
  Z_{\rm Potts}(Q=2) = \half (Z_2+Z_3+Z_4) = Z_{\rm Ising} \,,
\end{equation}
so we correctly find the Ising partition function.

\item The case $Q=1$. \\
  This case is {\it a priori} a bit intriguing. The partition function of the Potts
  model is then a trivial object (since there is only one state available for
  the whole lattice), while the general formulas for the central charge give
  in this particular case $c=-2$ ($Q=1$ so $\gamma=\pi/3$, $g=1/6, e_0=1/3$).
  This discrepancy occurs for the same reason as in
  the Berker-Kadanoff phase \cite{JS-PAF}: the level of the transfer matrix
  corresponding to a trivial partition function (and hence, formally, $c=0$)
  is very high in the spectrum, while the level generically dominating the
  thermodynamics (but which disappears {\it right at} $Q=1$ by quantum group
  truncation) corresponds to $c=-2$ (this means the free energy is a
  discontinuous function of $Q$ or of the boundary conditions \cite{JS-PAF}).

  Let us now see the mechanism in more details. The ground-state energy of our
  system in the untwisted case is twice the ground state energy of the {\af}
  XXZ model with $\Delta_0=-\cos2\gamma$. In the case $Q=1$ we have
  $\Delta_0=\half$.  The {\af} XXZ model with this value of the anisotropy
  is related 
  with the Potts model at $Q=1$ on the `non-physical self-dual line'
  \cite{SaleurPotts}. Recall that, meanwhile, the Potts model on the usual
  self-dual line is related to the {\af} XXZ chain at $2\Delta=-\sqrt{Q}$,
  so $\Delta=-\half$ in the case $Q=1$.

  Now we know that the energies of the {\af} XXZ at $\Delta=-\half$ are
  {\em minus} the energies of the {\af} XXZ at $\Delta=\half$ (this is the
  general mapping between $H_\Delta$ and $-H_{-\Delta}$). The ground-state
  energy of the {\af} XXZ at $\Delta=\half$ is the same, per unit length in
  the thermodynamic limit, as the one of the twisted {\af} XXZ, {\it i.e.} the
  ground-state energy of the percolation problem, {\it i.e.} $E_0=0$ in the
  proper normalisation. We thus conclude that the eigenvalue `corresponding to
  $Z=1$' in our spectrum is the {\it most excited among the subset of symmetric states}. 

  It is useful to see this mechanism at the level of partition functions as well. Start from \eqref{pgcd}
  and set $Q=1, \phi=\pi/3$. Then there is no contribution from the cross-geometry clusters. 
  Since $\cos 2\pi/3=\cos 4\pi/3 =-\half$, we have
  \begin{equation*}
    \sum_{m,m'} Z_{m,m'}(g) \cos(2\phi \ m \wedge m') =
    \frac{3}{2} \sum_{m=3p,m'=3p'}Z_{m,m'}- \half \sum_{m,m'} Z_{m,m'} \,.
  \end{equation*}
  Moreover, $m=3p$ is odd (resp. even) iff $p$ is odd (resp. even). Finally,
  $Z_{3p,3p'}(g)=Z_{p,p'}(9g)/3$. So we can rewrite
  \begin{equation}
    Z_{\rm Potts}(Q=1)=  \left(
      A \sum_{m,m' \ {\rm even}}
      + B \sum_{m \ {\rm even}, m' \ {\rm odd}}
      + C \sum_{m \ {\rm odd}, m' \ {\rm even}}
      + D \sum_{m,m' \ {\rm odd}}
    \right) \left[Z_{m,m'}(9g)-Z_{m,m'}(g)\right] \,.
  \end{equation}
  We can recombine terms using expressions for the $A,B,C,D$ in terms of the $Z_\nu$. We find
  \begin{eqnarray}
    Z_{\rm Potts}(Q=1) &=& \half(A-B-C-D)+(B-D) \sum_{m \  {\rm even}, m' }
    \left[ Z_{m,m'}(3/2) - Z_{m,m'}(1/6) \right] \nonumber \\
    &+& (C-D) \sum_{m , m' \ {\rm even}} \left[
      Z_{m,m'}(3/2)-Z_{m,m'}(1/6)
    \right] \,,
  \end{eqnarray}
  where we have specialized to $g=1/6$ and used  Euler's identity:
  \begin{equation}
    \sum_{m,m'~even} Z_{m,m'}(3/2)-Z_{m,m'}(1/6) =
    \half \left[Z_c(6)-Z_c((2/3)\right] = 1 \,.
  \end{equation}
  Now we have 
  \begin{equation}
    \sum_{m \ {\rm even}, m' } \left[ Z_{m,m'}(3/2) - Z_{m,m'}(1/6) \right]
    =\sum_{m\ {\rm even},e} \left[ Z_{em}(3/2) - Z_{em}(1/6) \right]
  \end{equation}
  and
  \begin{equation}
    \sum_{m , m' \ {\rm even}} \left[ Z_{m,m'}(3/2)-Z_{m,m'}(1/6) \right]
    =\half \sum_{m,e\ {\rm half-integer}} \left[ Z_{em}(3/2)-Z_{em}(1/6) \right] \,.
  \end{equation}
  Both terms can be shown to vanish exactly. We conclude that
  \begin{equation}
    Z_{\rm Potts}(Q=1)=0 \,.
  \end{equation}
  This means there are exact cancellations among states in the low-energy
  spectrum: the (unique) state that would correspond to the trivial partition
  function is very highly excited and does not contribute to the conformal partition
  function (at $c=-2$ in this case). 
\end{itemize}

\end{document}